\documentclass[10pt,letterpaper]{article}
\usepackage[top=0.85in,left=2.75in,footskip=0.75in]{geometry}

\usepackage{amsmath,amssymb}

\usepackage{changepage}

\usepackage[utf8x]{inputenc}

\usepackage{textcomp,marvosym}

\usepackage{cite}

\usepackage{nameref,hyperref}

\usepackage{microtype}
\DisableLigatures[f]{encoding = *, family = * }

\usepackage[table]{xcolor}

\usepackage{array}

\newcolumntype{+}{!{\vrule width 2pt}}

\newlength\savedwidth

\raggedright
\usepackage[aboveskip=1pt,labelfont=bf,labelsep=period,justification=raggedright,singlelinecheck=off]{caption}

\makeatletter
\renewcommand{\@biblabel}[1]{\quad#1.}
\makeatother

\usepackage{lastpage,fancyhdr,graphicx}
\usepackage{epstopdf}
\fancyheadoffset[L]{2.25in}
\fancyfootoffset[L]{2.25in}
\newcommand{\Real}{\mathbb{R}}

\newtheorem{prop}{Proposition}

\begin{document}
\vspace*{0.35in}

\begin{flushleft}
{\Large
 \textbf\newline{Simple model of complex dynamics of activity patterns in
developing networks of neuronal cultures} }
\newline
\\
Ivan  Y. Tyukin\textsuperscript{1,3,4,*},
Dmitriy Iudin\textsuperscript{1,2},
Feodor Iudin\textsuperscript{1},
Tatiana Tyukina\textsuperscript{4},
Victor Kazantsev\textsuperscript{1,2},
Irina Muhina\textsuperscript{1},
Alexander N. Gorban\textsuperscript{1,4}
\\
\bigskip
\bf{1} Nizhny Novgorod State University, 23 Gagarin ave., 603950, Nizhny Novgorod, Russia
\\
\bf{2} Institute of Applied Physics of RAS, 46 Uljanov str., 603950, Nizhny Novgorod, Russia
\\
\bf{3} Saint-Petersburg State Electrotechnical University (LETI), prof. Popova str. 5, 197376, Saint-Petersburg, Russia
\\
\bf{4} University of Leicester, Leicester LE1 7RH, United Kingdom
\\
\bigskip

* I.Tyukin@le.ac.uk

\end{flushleft}
\section*{Abstract}
Living neuronal networks in dissociated neuronal cultures are widely known for their ability to generate highly robust spatiotemporal activity patterns in various experimental conditions. These include neuronal avalanches satisfying the power scaling law and thereby exemplifying self-organized criticality in living systems. A crucial question is how these patterns can be explained and modeled in a way that is biologically meaningful, mathematically tractable and yet broad enough to account for neuronal heterogeneity and complexity. Here we propose a simple model which may offer an answer to this question. Our derivations are based on just few phenomenological observations concerning input-output behavior of an isolated neuron. A distinctive feature of the model is that at the simplest level of description it comprises of only two variables, a network activity variable and an exogenous variable corresponding to energy needed to sustain the activity and modulate the efficacy of signal transmission. Strikingly, this simple model is already capable of explaining emergence of network spikes and bursts in developing neuronal cultures. The model behavior and predictions are supported by empirical observations and published experimental evidence on cultured neurons behavior exposed to oxygen and energy deprivation.   At the larger, network scale, introduction of the energy-dependent regulatory mechanism enables the network to balance on the edge of the network percolation transition. Network activity in this state shows population bursts satisfying the scaling avalanche conditions. This network state is self-sustainable and represents a balance between global network-wide processes and spontaneous activity of individual elements.

\section*{Author Summary}
We propose a model of spiking activity in living developing neuronal networks in dissociated neuronal cultures. Such neuronal cultures generate a broad range of complex yet highly robust
spatiotemporal activity patterns in various experimental conditions. These  patterns are often regarded as neuronal avalanches that satisfy the power scaling law and thereby exemplify self-organized criticality in living systems. In this work we approach an important question of how these patterns can be explained and modeled by just few simple equations and variables of which the phenomenological description make clear biological and physical sense. A distinctive feature of the model is that at the simplest level of description it comprises of only two variables, the network activity variable and an exogenous variable corresponding to energy needed to sustain the activity and modulate the efficacy of signal transmission. Despite its apparent simplicity the model is capable of explaining several common behavioral features of developing neuronal cultures, including emergence of network spikes and bursts as a function of days in development. We show that when the model is extended to the larger, network scale, introduction of the energy regulatory mechanism allows the system to balance on the edge of the network percolation transition. Network activity in this state shows population bursts satisfying the scaling avalanche conditions. This network state is self-sustainable and represents a balance between global network-wide processes and spontaneous activity of individual elements.


\section*{Introduction}
Exploiting physics' concepts in life sciences has long been reconginzed as a successful strategy for developing systematic understanding of complex phenomena observed in empirical data. One of the most striking and fashionable illustrations facilitating potential and power of this approach is the well-known concept of self-organized criticality (SOC) -- the ability of systems to self-tune to a critical state. Initially proposed as a model for explaining how an abstract system can remain at a critical state in presence  of perturbations \cite{Bak:97, Iudin:13}, the
concept is now broadly used for describing biological neural networks (see e.g. \cite{Christensen:98, Bornholdt:00}). It has been shown in \cite{Iudin:15}, \cite{Levina:2007} that networks whose structural evolution is linked with the node's dynamcis  can exhibit highly robust global SOC-like behavior. What is important is that this behavior can be maintained by simple local adjustment rules.

Examples of SOC-like behavior have been found in experimental studies of neuronal cultures
\cite{beggs:03,beggs:04,Chiappalone:08}. The cultures grow autonomously and form synaptically coupled networks of living cells. After a period of initial growth and development the cultures start to generate spontaneous activity patterns in the form of population bursts. These bursts were shown to satisfy the power scaling law and hence are often referred as neuronal avalanches \cite{beggs:03, beggs:04}.

Since then a number of mathematical models have been proposed to capture and analyse mechanisms involved in the generaaion of  spontaneous burst of activity in neuronal networks. The spectrum of network's features linked to
emergence of persistent bursts includes, but is not limited to, e.g. re-wiring, delays \cite{gong:04,gong:07}, frequency dependent and spike-timing dependent synaptic plasticity \cite{Markram:00,Izhikevich:04,Izhikevich:06}. With regards to the mathematical frameworks describing neuronal avalanches, models of network's growth \cite{Abott:07} and stochastic networks \cite{Cowan:10} have been proposed.

It has been shown in \cite{Masquelier:13} (see also references therein) that population spikes and bursts can be attributed to cell's adaptation and short time plasticity mechanisms. The authors showed that population spikes similar to the ones observed in in-vitro cultures can occur in networks of excitatory model neurons with leaky integrate and fire dynamics. These neurons where subjected to Gaussian white noise and equipped with adaptation and short term plasticity mechanism. The network connectivity was all-to-all. The influence of neuronal connectivity on bursts was studied in \cite{Gritsun:12}. It has been found that the number of synaptic connections per neuron, may play an important role for spiking and bursting activity in cultures. Additional links between connectivity development, firing activity homeostasis, and criticality are exposed in \cite{Tetzlaff:10}.

In this work we further contribute to the idea that several features of complex and critical behavior (e.g. the neuronal avalanches, super-bursts, periodic and chaotic spiking) observed in live neuronal cultures and networks can be explained by just few variables. These variables can be linked to  local connectivity patterns (expressed e.g. by connection densities between cells) and neuronal activation dynamics.

We demonstrate that main critical transitions can be captured by a hierarchy of simple models.
%
%
Starting from elementary phenomenological description of neural firing, we present a simple $2D$ mean-field model that is capable of a broad range of behaviors conistent with the ones observed in cultures. Remarkably, the network connectivity parameter appears to be a natural bifurcation parameter of the model; it regulates emergence of activity spikes from an initially silent mode to occasional network bursts which, in turn, eventually develop into a whole-network activation. This is in good agreement with previous works on SOC phenomena in developing neuronal networks \cite{Levina:2007}, \cite{Tetzlaff:10}. In contrast to \cite{Levina:2007}, \cite{Tetzlaff:10}, however, we consider the problem from a different angle. Instead of focusing on the activity-connectivity interplay we consider and analyze the system's dynamics in the activity-energy plane for various values of connectivity. This allows additional modelling capabilities, including e.g. the analysis of how oxygen and energy deprivation affects the network's behavior and dynamics.  Moving further to multi-agent model reveals emergence of neuronal avalanches showing  power  law scaling.

The manuscript is organized as follows. In ~\nameref{sec:results} we present ingredients of the model at three different levels of phenomenological detail. We begin with a simple percolation-based geometric model describing the evolution of cells' connectivity is presented. The model allows to accommodate biologically relevant features such as axons and dendrites; it also enables to replicate directional connectivity that is inherent to living systems including neuronal cultures. The model analysis reveals that sharp changes in the overall clustering and connectivity of the evolving network in both directed and undirected settings is determined by a single parameter describing average connection density in the network. The analysis is followed by expressing dynamics of neuronal activity by a mean-field approximation. We show that the corresponding single-dimensional model does not explain network spikes and bursts frequently observed in developing cultures. This limitation, however, can be resolved if neural activation is linked to an additional exogenous regulatory energy variable. Introduction of the latter  variable needs an additional comment. It behaves as a sort of ``energy'' or a resource. Its physical nature, nevertheless, may or may not be associated with a specific type of the physical energy. A prototype of such energy, the  notion of {\it adaptation energy}, was introduced by Selye in his analysis of physiological adaption \cite{Selye:38a}, \cite{Selye:38b} and was successfully employed in modelling of various complex phenomena \cite{Gorban:10}, \cite{Gorban:15}.
%
%
%
%
%
%
%
%
The extended model is shown to be able to reproduce periodic spiking, irregular dynamics, and population bursts observed in live cultures. What is important is that dynamic regimes exhibited by the model can be regulated by just a single parameter corresponding to network connectivity.  Next we provide results of large-scale simulation of evolving network of agents of which the activation probability depends on their current energy level. The network dynamics in this state shows population bursts satisfying the scaling avalanche conditions. This network state is self-sustainable and represents energetic balance between global network-wide processes and spontaneous activity of individual elements. The results are compared with empirical data. Section ~\nameref{sec:methods} describes details of electrophysiological measurements and experiments, and Section~\nameref{sec:conclusion} concludes the paper.

\section*{Results}\label{sec:results}

\subsection*{Model}

\paragraph{Geometric model}\label{sec:geometry}
We start with geometrical arrangement of the network elements. Consider a network of $N$ neurons whose spatial coordinates are randomly and uniformly distributed in the unit square. Each individual neuron is described by two basic elements. The first is the region of reception of inputs represented by a circle of a given radius $R$. The circle models neuron's ability to sense input signals from other neurons, and is referred to as the dendrite region (in biology, dendrite is an input). The second element is an axon (in biology, output), which in our model is simulated by a straight segment of length $H$ (on the mature stage of the network development $H > R$) and whose end point is acting as a transmitter of the neuron's signal. If this point reaches out to the dendrite region of another neuron, a connection is established between these neurons \cite{Abott:07}. There are three different ways that yield geometrical coupling or connectivity of the network elements:

\noindent
{\it Case 1}: cells without axons, i.e. $H=0$. In this case $N$ circles with radius $R$  are randomly and uniformly distributed in the unit square. If a circle {\bf A}  overlaps with a circle {\bf B}, and circle {\bf B} is connected with a circle {\bf C}, then {\bf A} is connected with {\bf C}. Thus, a path between two distant cells can be defined as a chain of overlapping circles joining these cells. Emergence of large groups of connected elements in this network can be analyzed within the framework of standard circle percolation problem. Let $n$ be the cells density defined e.g. as the number of circles' centers in a unit area. According to \cite{Ziff:07,Kirkpatrick:71}, emergence of large groups of interconnected cells, the percolation transition, in a set of randomly distributed circles can be characterised by the mean number of centers that fall within a circle of radius $R$:
\begin{equation}\label{eq:B2D}
B=\pi R^2n.
\end{equation}
In particular, there exists a critical concentration $B=B_c$ at which two arbitrary circles become connected with high probability. Thus percolation occurs and a large cluster of connected circles
appears. In contrast with typical thermal phase transitions where a transition between two phases occurs at a critical temperature, percolation transition relates to distribution and topology of clusters corresponding to the values of $B$ in a neighborhood of $B_{c}$. At low values of $B$ only small clusters of overlapping circles exist. When the concentration $B$ increases the average size of the clusters increases too. At the critical concentration value, $B=B_{c}$, a large cluster
appears; it includes groups of cells that are close to the opposing boundaries of the original square. This cluster is called \emph{spanning cluster} or percolating cluster. In the thermodynamic limit, i.e. in the infinite size system limit the spanning cluster is called \emph{infinite cluster}. For scalar problem the value of $B_c \approx 1.1$.

\noindent
{\it Case 2}. Cells have axons, $H>0$, and axons are allowed to transmit signals in both directions. Each neuron can be represented as an undirected pair of head- and tail-circles both having radius $R$. When the head-circle or the tail-circle of an neuron overlaps with the head- or the tail-circles
of another neuron we consider these neurons connected. Despite obvious difference of this setting from the previously considered one in that we are to operate with dipoles here rather than with just circles as in Case 1, the problem remains within a class of scalar percolation, albeit for dipoles of circles not just a single circle.

\noindent
{\it Case 3}. Cells have  axons, $H>0$, and these axon can only transmit signals along a straight line that determines direction of connectivity for a given cell. The coupling direction from soma to synaptic terminal has isotropic distribution, and hence each neuron could be represented as a directed pair of head- and tail-circles both having radius $R$. Vectors linking centers of the head- and tail-circles are allowed to have arbitrary direction. Their lengths, $H$, however, are fixed. When the tail-circle of a neuron overlaps with a head-circle of another neuron the pair is considered as connected. In contrast with two other ways of establishing neuronal connectivity considered above this is the most realistic scenario. It is no longer within the scope of simple scalar circle percolation framework but is a vector percolation problem.

\begin{figure}[htb]
\centering\noindent
\includegraphics[width=0.7\textwidth]{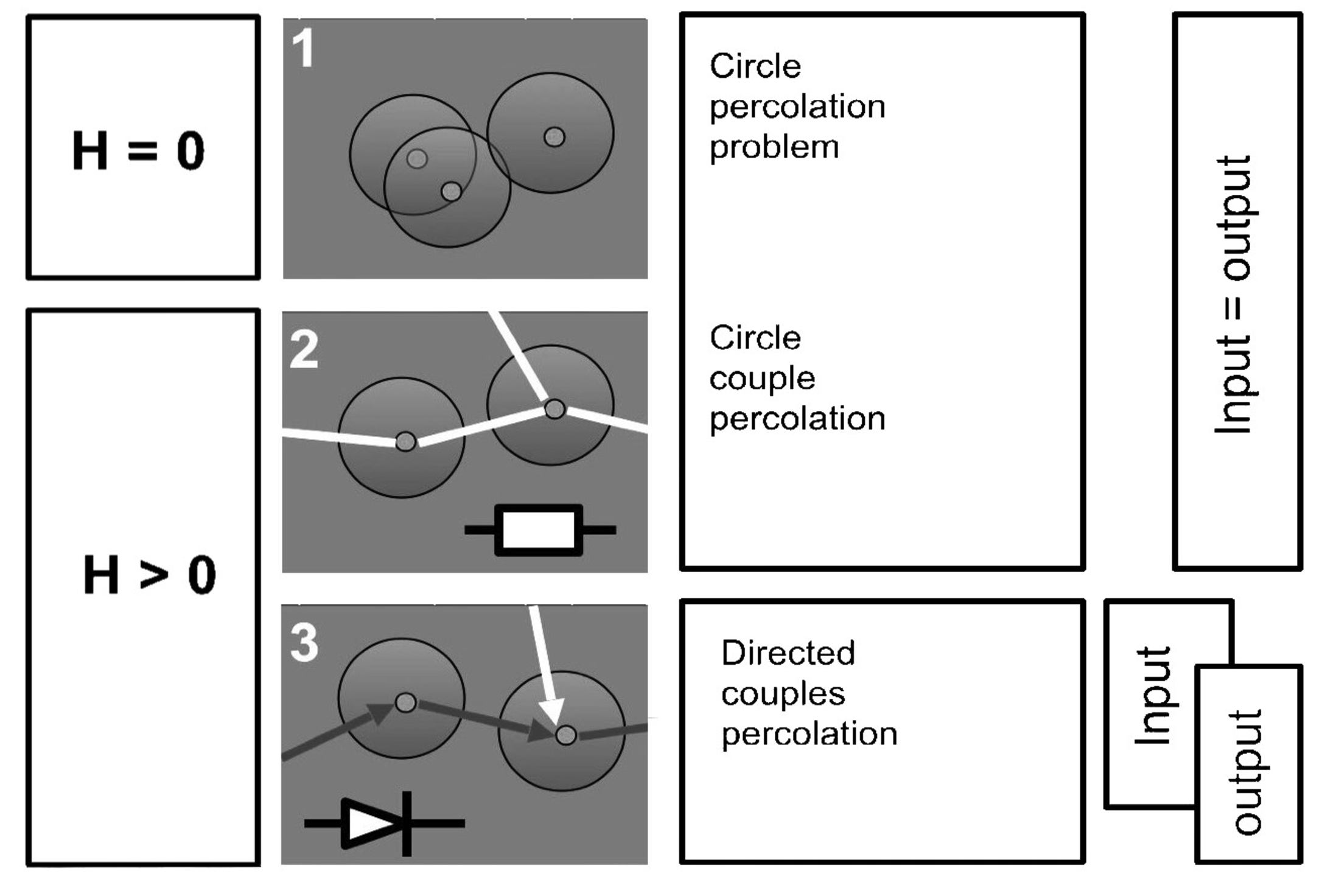}
\caption{Schematic representation of three different percolation settings for the geometrical model.}
\label{fig:scheme1}
\end{figure}

\begin{figure}[htb]
\centering\noindent
\includegraphics[width=0.5\textwidth]{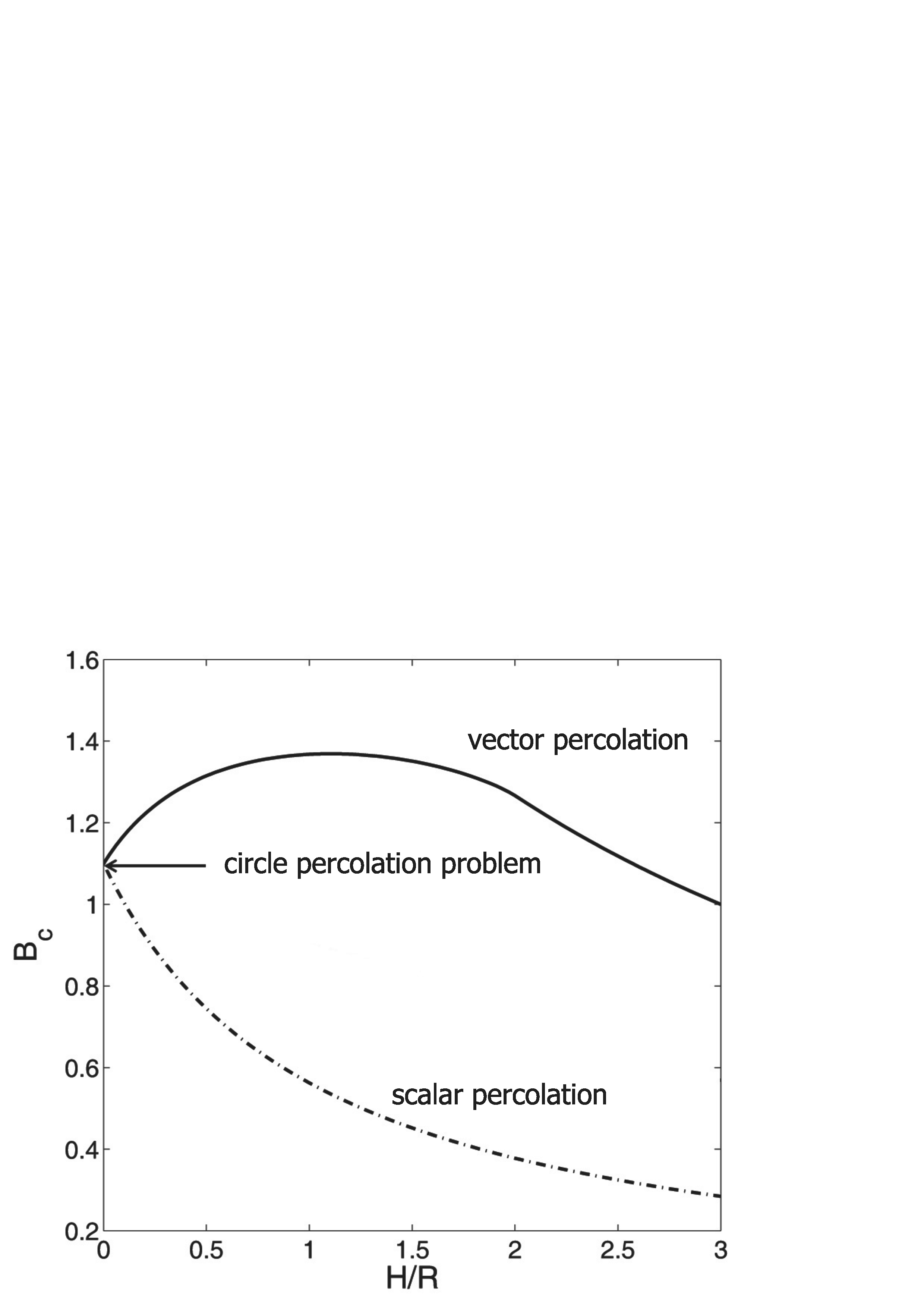}
\caption{Dependence of the percolation threshold parameter, $B_{c}$, on $H/R$ for 2D scalar and vector percolation problems (Case 2 and 3 respectively).}
\label{fig:thresh}
\end{figure}

The three cases are illustrated with Fig.~\ref{fig:scheme1}. Fig.~\ref{fig:thresh} shows dependence of the percolation threshold parameter, $B_c$, on the ratio $H/R$.
In accordance with he definition of $B$ in (\ref{eq:B2D}) the cell's concentration variable $B$ can be related to the expected number of neurons that are connected to a randomly chosen query neuron in the system. The latter quantity is known as the average network coordination number $z$. As can be seen from Fig.~\ref{fig:thresh} the value of the coordination number $z$ that corresponds to $B_c$ is always bounded from above. $B_c\leq 1.4$ for all configurations considered so far.

The above analysis reveals that  networks with coordination numbers exceeding these critical values are likely to form a spanning cluster that is capable of connecting opposite edges of the system \cite{Iudin:12, Iudin2:15}. Thus, intuitively, one can argue that signals initiated by spontaneous activation of neurons in the cluster can spark waves of activity through the whole network. Such conclusion is based on the restrictive assumption that neurons always elicit spikes in response to a spike on their input. Not only this assumption does not necessarily hold true, but also the above geometrical model alone does not explain the wealth of excitation propagation phenomena observed in cultures. To account for more plausible situations as well as to possibly increase explanatory power of the model two additional variables are introduced: one is the maximal probability $p$ of neuronal
activation in response to incoming spike, and the other is an exogenous ``resource'' variable $E$ determining if a neuron has enough energy to elicit a spike. The consequences of adding these two variables are discussed in the next sections.

\paragraph{Mean-field dynamic model of neuronal excitation}\label{sec:mean_field_approximation}
We begin  with a simple mean-field approximation of the dynamics of neural excitation in  the system. Consider a connected network of neuronal cells. Let $z$ be the expected coordination number,
i.e. the expected number of neighbors of a randomly chosen query cell. Suppose that at a time instance $t$ some neurons in the network are excited. Let $q_t$ denote the number of these neurons
relative to the total number of cells the whole network. If the value of $z$ is sufficiently large then the number of excited neurons among all neighbors of a given neuron can be estimated as
$zq_t $. Let $p$ be the probability of neuronal activation in response to activation of at least one neuron from its nearest neighbours. Moreover, we suppose that all excitatory signals are independent. Thus the probability that a given neuron is activated equals to $ 1-(1-p) ^ {zq_t}$, and hence the expected proportion of all excited neurons at the time step $t+1$ is:
\begin{equation} \label{eq:1.41}
q_{t+1}=1-(1-p) ^ {zq_t}.
\end{equation}
The range of dynamics which model (\ref{eq:1.41}) is capable to reproduce is summarized in Proposition \ref{prop:no_spikes}.
\begin{prop}\label{prop:no_spikes} Consider (\ref{eq:1.41}) and suppose that $p\in(0,1)$, $z>0$ be constants. Then the interval $[0,1]$ is forward-invariant, all forward orbits $q_t$ are monotone functions $t$, and the point map (\ref {eq:1.41}) has only fixed points as attractors. Furthermore
\begin{enumerate}
\item if $-z\log(1-p)\leq 1$ then the map has only one fixed point, $q^\ast_1=0$, and it is an attractor;
\item if $-z\log(1-p)> 1$ then the map has only two fixed points with $q^\ast_1=0$ being a repeller and the other one, $q^\ast_2\in(0,1)$, a stable attractor.
\end{enumerate}
\end{prop}
Proof of Proposition \ref{prop:no_spikes} is provided in the Appendix.

According to Proposition \ref{prop:no_spikes}, asymptotic mean-field dynamics (\ref{eq:1.43}) of the network is a steady activity at an equilibrium in the entire domain of the model's feasible parameters: $p\in(0,1)$, $z>0$. Moreover, all transients are monotone trajectories. Emergence of the unique non-zero asymptotic activity is fully determined by the values of the connectivity parameter, $z$, and the probability of neural activation, $p$. Critical values of these parameters, e.g. the critical connectivity, $\overline{z}_1$, at which the transition occurs, satisfies:
\begin{equation}\label {eq:1.43}
\overline{z}_1=\frac{-1}{\log(1-p)}.
\end{equation}
For $\overline{z}_1\gg1$, this relation is approximately reciprocal: $p\simeq 1/\overline{z}_1$. Indeed, expressing $p=1-\exp\left(-\frac{1}{\overline{z}_1}\right)$ from (\ref{eq:1.43}), and expanding $\exp\left(-\frac{1}{\overline{z}_1}\right)$ as a power series with respect to $1/\overline{z}_1$ one obtains:
\[
p=\frac{1}{\overline{z}_1}+O\left(\frac{1}{{\overline{z}_1}^{2}}\right).
\]

The value of the stable equilibrium, $q^{\ast}_2$, can be estimated as follows. At the steady state we have that $1- q^{\ast}_2 = (1-p)^{z q^\ast_2}$. Hence $(1-q^{\ast}_2)^{1/ q^{\ast}_2}=(1-p)^z$. It is well known that  $(1-x)e^{-1}<(1-x)^{1/x}$ for all $x\in (0,1)$. Thus
\[
(1- q^{\ast}_2) < e(1-p)^z \Rightarrow  q^{\ast}_2 > 1 - e(1-p)^z.
\]
Observe that the larger is the value of the  connectivity parameter, $z$, the higher is the level of the mean-field network activity.

Whilst model (\ref{eq:1.43}) is consistent with the very basic observation that increasing the network's overall connectivity may lead to emergence of  a self-sustained activity in the network, the model's explanatory capability is limited. The model does not explain widely-reported richness of the dynamics in live neuronal cultures, including emergence of spontaneous activity bursts and irregular and seemingly chaotic spikes.
%
%
%
%

This limitation is not surprising since (\ref{eq:1.41}) is a crude approximation of the network's dynamics. Model (\ref{eq:1.41}) does not account for a broad spectrum of biological mechanisms involved in spike generation and assumes that the neuron's ability to produce spikes depends exclusively on stimulation. A possible way to overcome this unrealistic assumption is to explicitly account for these missing mechanisms. To keep the model simple, we account for joint effect of these mechanisms by adding a single energy-like variable $E_t$ to (\ref{eq:1.41}). The new variable determines the neuron/network's ability to produce a spike depending on the amount of resources or ``energy'' available. Generic models of this type have been proposed and analyzed in \cite{Gorban:09, Gorban:10} in the context of adaption to stress and external environmental factors. These models have been shown to capture periodic and irregular behavior in multi-agent systems \cite{Gorban:15}.

\begin{figure}
\centering\noindent
\includegraphics[width=.5\textwidth]{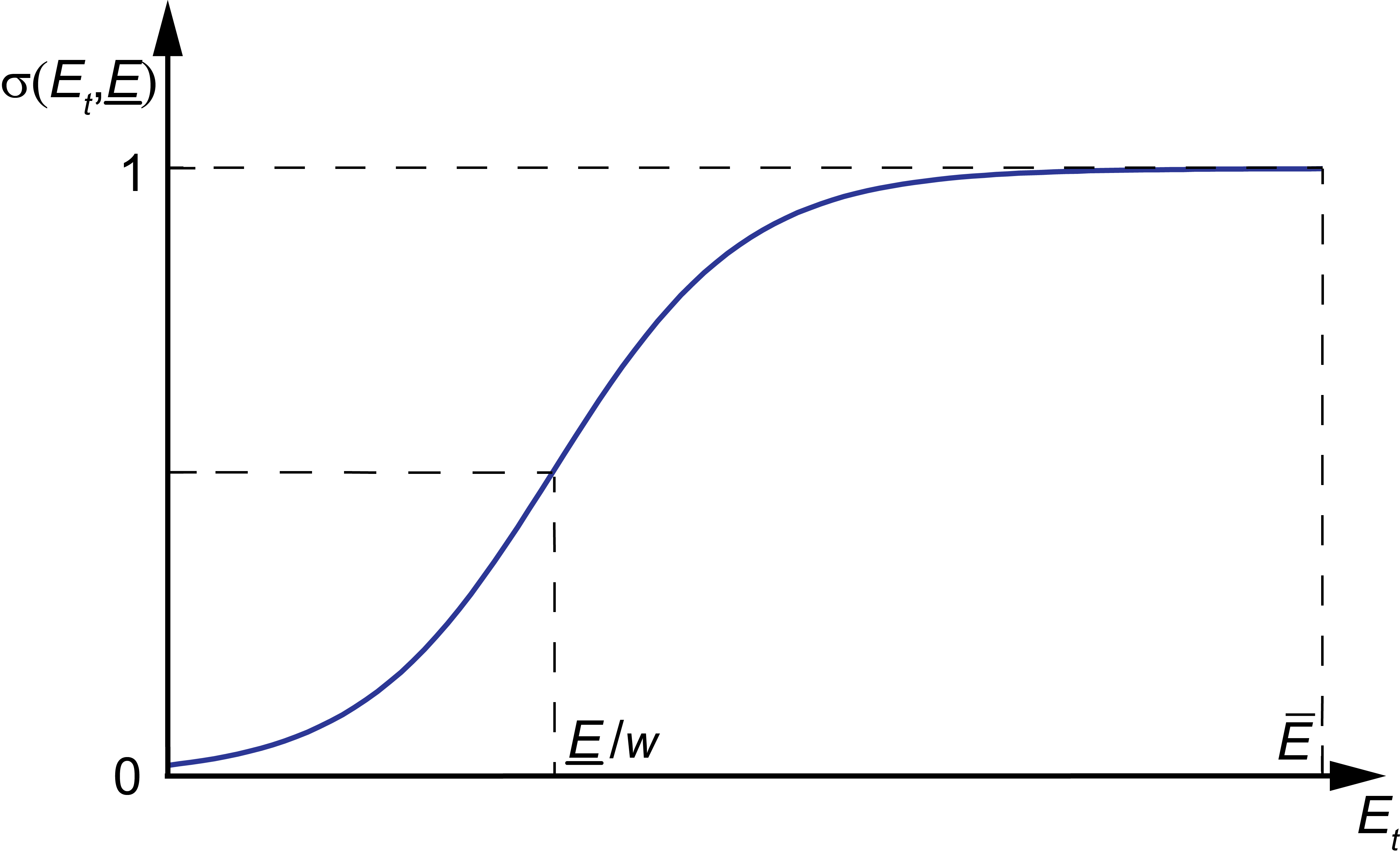}
\caption{General shape of the function $\sigma(\cdot,\underline{E})$.}\label{fig:sigma}
\end{figure}

Here we extend the original phenomenological mean-field model (\ref{eq:1.41}) as follows:
\begin{equation}\label{eq:mean_field_adaptive}
\begin{split}
q_{t+1}=&1-(1-p\cdot \sigma(E_t,\underline{E})) ^ {zq_t}\\
E_{t+1}=& (1-\varepsilon)E_t + \varepsilon \overline{E} - r q_t\mathcal{H}\left(E_t  - r q_t\right),
\end{split}
\end{equation}
where
\[
\sigma(E_t,\underline{E})=\frac{1}{2}(\tanh(w E_t - \underline{E})+1),
\]
and $\mathcal{H}$ is the Heaviside function. In (\ref{eq:mean_field_adaptive}) $p$ is the maximal probability of neuronal activation, $z>0$ is the coordination number, $E_t$ is the exogenous phenomenological ``energy resource'' variable; $r>0$ is the energy cost of neuronal activation, $\underline{E}>0$ and $w>0$ are parameters that determine the minimal activation probability and the energy activation threshold, $\overline{E}>\underline{E}$ is the energy recovery value, $\varepsilon\in(0,1)$ is the energy relaxation parameter. General shape of the function $\sigma(\cdot,\underline{E})$ in the energy-dependent synaptic efficacy component, $p\sigma(\cdot,\underline{E})$, is shown in Fig.~\ref{fig:sigma}.

Mean-field model (\ref{eq:mean_field_adaptive}) of the network dynamics inherits phenomenological transparency of (\ref{eq:1.43}). It does, however, account for generic constraints of spike-generation through the new energy variable $E_t$ and energy-modulated synaptic efficacy $p\sigma(\cdot,\underline{E})$.  Despite retaining simplicity, the model produces remarkably rich dynamics. Its equilibria, however, can still be described by just few parameters as follows from Proposition \ref{prop:energy_model} below.

\begin{prop}\label{prop:energy_model} Consider (\ref{eq:mean_field_adaptive}) with $p\in(0,1)$. Then the domain
\[
\{(q,E) \ | q\in [0,1], E\in [0, \overline{E}]\}
\]
is forward-invariant. In addition,
\begin{enumerate}
\item If  $-z \log\left(1-p\sigma\left(\overline{E},\underline{E}\right)\right)\leq 1$
then (\ref{eq:mean_field_adaptive}) has only one fixed point:
\begin{equation}\label{eq:fixed_point_1}
 (q^\ast_1,E^\ast_1)=(0,\overline{E}).
\end{equation}
This fixed point is an attractor when $-z \log\left(1-p\sigma\left(\overline{E},\underline{E}\right)\right)< 1$.

\item If $-z \log\left(1-p\sigma\left(\overline{E},\underline{E}\right)\right)> 1$ and ${\overline{E}}/{r}\geq 1$
then (\ref{eq:mean_field_adaptive}) has at most two fixed points:  fixed point (\ref{eq:fixed_point_1}) and, if exists, an additional fixed point $(q^\ast_2,E^\ast_2)$:
\begin{equation}\label{eq:fixed_point_2}
\begin{split}
 &(q^\ast_2,E^\ast_2),  \   0<q_2^\ast\leq \frac{\overline{E}}{r\left(1+\frac{1}{\varepsilon}\right)}, \ q_2^\ast<1, \\
 & q_2^\ast= 1 - \left(1-p \sigma\left(\overline{E}-\frac{r q^\ast_2}{\varepsilon},\underline{E}\right)\right)^{q_2^\ast z},  \ E^\ast_2=\overline{E}- \frac{r q_2^\ast}{\varepsilon}.
\end{split}
\end{equation}
The fixed point $(q^\ast_1,E^\ast_1)$ a repeller.

\item If $-z \log\left(1-p\sigma\left(\overline{E},\underline{E}\right)\right)> 1$ and ${\overline{E}}/{r}<1$ then (\ref{eq:mean_field_adaptive}) has at most three fixed points. The fixed point $(q^\ast_1,E^\ast_1)$ specified by (\ref{eq:fixed_point_1}) and, possibly, additional two fixed points: the fixed point $(q^\ast_2,E^\ast_2)$ specified by (\ref{eq:fixed_point_2}) and a fixed point
\begin{equation}\label{eq:fixed_point_3}
\begin{split}
&(q^\ast_3, E^\ast_3), \ 1> q^\ast_3>\frac{\overline{E}}{r},  \\
& q^\ast_3=1-(1-p\sigma(\overline{E},\underline{E}))^{q^\ast_3 z}, \ E^\ast_3=\overline{E}.
\end{split}
\end{equation}
The fixed point $(q^\ast_1,E^\ast_1)$ is a repeller, and $(q^\ast_3, E^\ast_3)$, if exists, is a stable attractor.
\end{enumerate}
\end{prop}
Proof of Proposition \ref{prop:energy_model} is provided in the Appendix.

\begin{figure}
\centering
\includegraphics[width=0.6\textwidth]{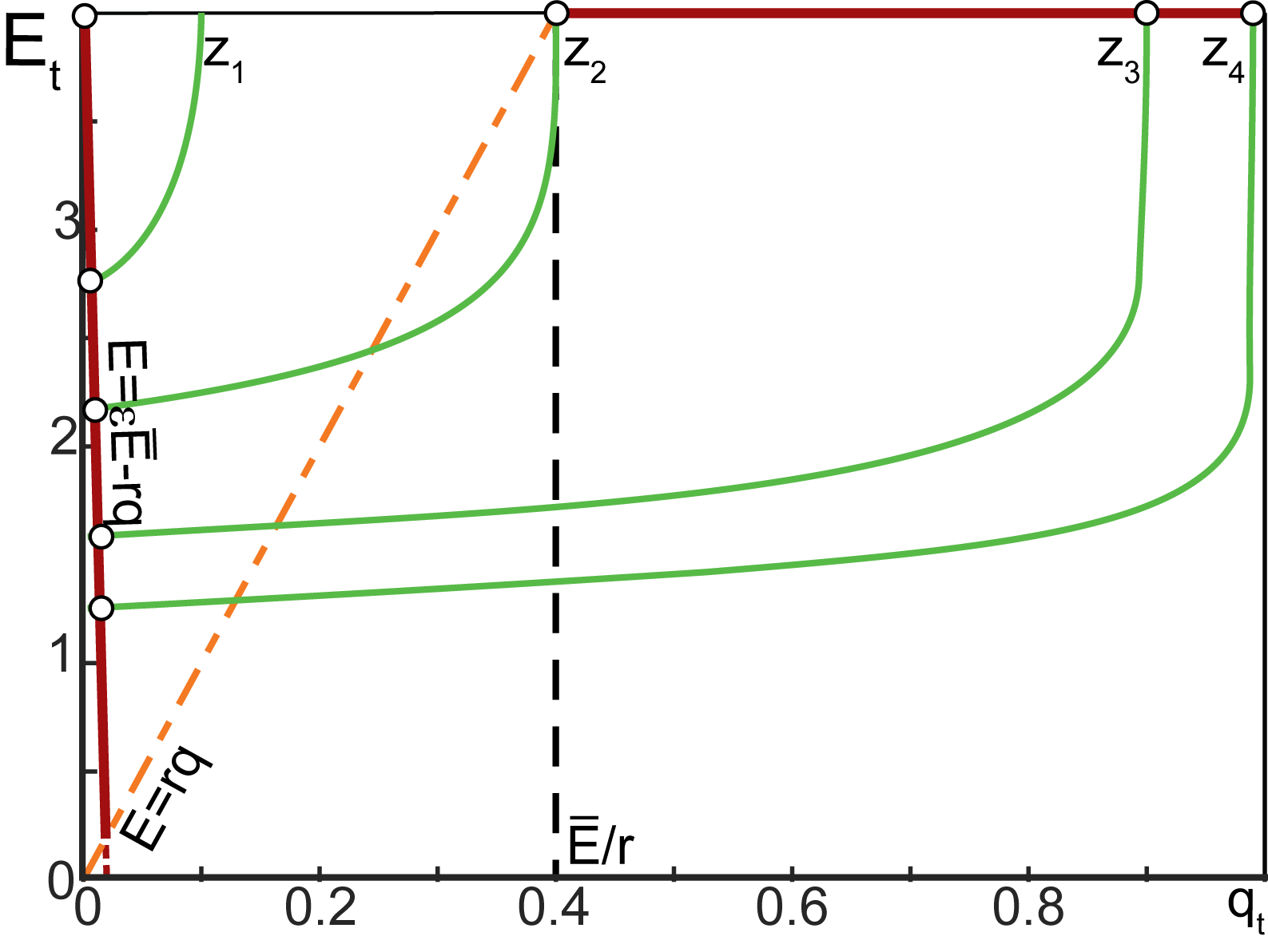}
\caption{Arrangement of equilibria in (\ref{eq:mean_field_adaptive}). The  model parameters were set as follows: $w=1.5$, $p=0.1$, $\overline{E}=4$, $\underline{E}=2$, $\varepsilon=0.05$, $r=10$, with $z$ taking values in the set $\{10,12.2,25,50\}$ (these values are denoted as $z_1$, $z_2$, $z_3$ and $z_4$).}\label{fig:critical_points}
\end{figure}

An illustration showing relationships between parameters of the model and emergence of the three different equilibria described in Proposition \ref{prop:energy_model} is provided in Fig. \ref{fig:critical_points}. The equilibria are shown as white circles.  Green lines show curves
\begin{equation}\label{eq:sigma_inv}
E^\ast=\sigma^{-1}(q^\ast,\underline{E})= \frac{1}{w}\tanh^{-1}\left(\frac{2}{p}\left(1-(1-q^\ast)^{\frac{1}{q^\ast z}}\right)-1\right)+\frac{\underline{E}}{w}
\end{equation}
as functions of $q^\ast>0$. According to the first equation of (\ref{eq:mean_field_adaptive}), all equilibria of the model with $q^\ast\neq 0$ must belong to these curves (see also the proof of Proposition \ref{prop:energy_model} in Appendix). Depending on the value of $z$, the curves move up and down, and intersect with line segments (shown as red solid lines in Fig. \ref{fig:critical_points}):
\[
E^\ast=\overline{E}-r \frac{q^\ast}{\varepsilon}, \ 0<q^\ast\leq \frac{\overline{E}}{r\left(1+\frac{1}{\varepsilon}\right)}, \ q^\ast<1
\]
and
\[
E^\ast=\overline{E}, \ \frac{\overline{E}}{r}<q^\ast<1.
\]
These intersections correspond to equilibria (\ref{eq:fixed_point_2}) and (\ref{eq:fixed_point_3}), respectively.
The equilibria persist over intervals of $z$, and the greatest lower bounds of these intervals (critical values of $z$) are:
\begin{equation}\label{eq:critical}
\overline{z}_2 = \frac{-1}{\log(1-p\sigma(\overline{E},\underline{E}))} \ \mbox{and} \
\ \overline{z}_3 = \frac{\log\left(1-\frac{\overline{E}}{r}\right)}{\frac{\overline{E}}{r} \log (1-p\sigma(\overline{E},\underline{E}))}, \ \frac{\overline{E}}{r}<1.
\end{equation}
We also note (see the proof of Proposition \ref{prop:energy_model}) that equilibria (\ref{eq:fixed_point_2}) are always above or on the line $E=r q$ (dashed orange line in Fig. \ref{fig:critical_points}), whereas equilibria (\ref{eq:fixed_point_3}) are to be below this line.

According to Propositions \ref{prop:no_spikes}, \ref{prop:energy_model},  models (\ref{eq:1.41}) and  (\ref{eq:mean_field_adaptive}) share some similarity.  For $z$ sufficiently small all orbits are attracted to a single equilibrium. At this equilibrium, the systems are silent. When $z$ increases and eventually exceeds the first critical value  (eq. (\ref{eq:1.43}) for (\ref{eq:1.41})  and $\overline{z}_2$ for (\ref{eq:mean_field_adaptive})), the silent equilibrium becomes a repeller and the systems start to exhibit non-zero activity. However, further increases of $z$ trigger drastically different dynamics in these models.

\begin{figure}
\centering
\includegraphics[width=\textwidth]{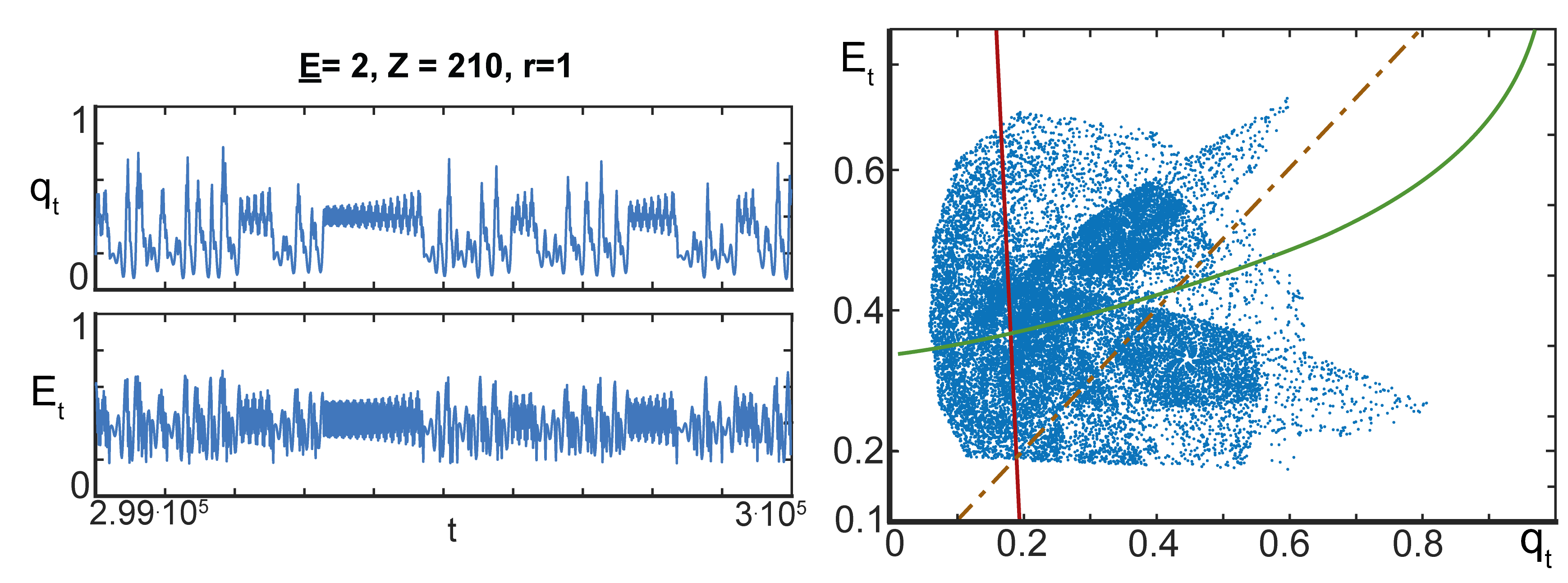}
\includegraphics[width=\textwidth]{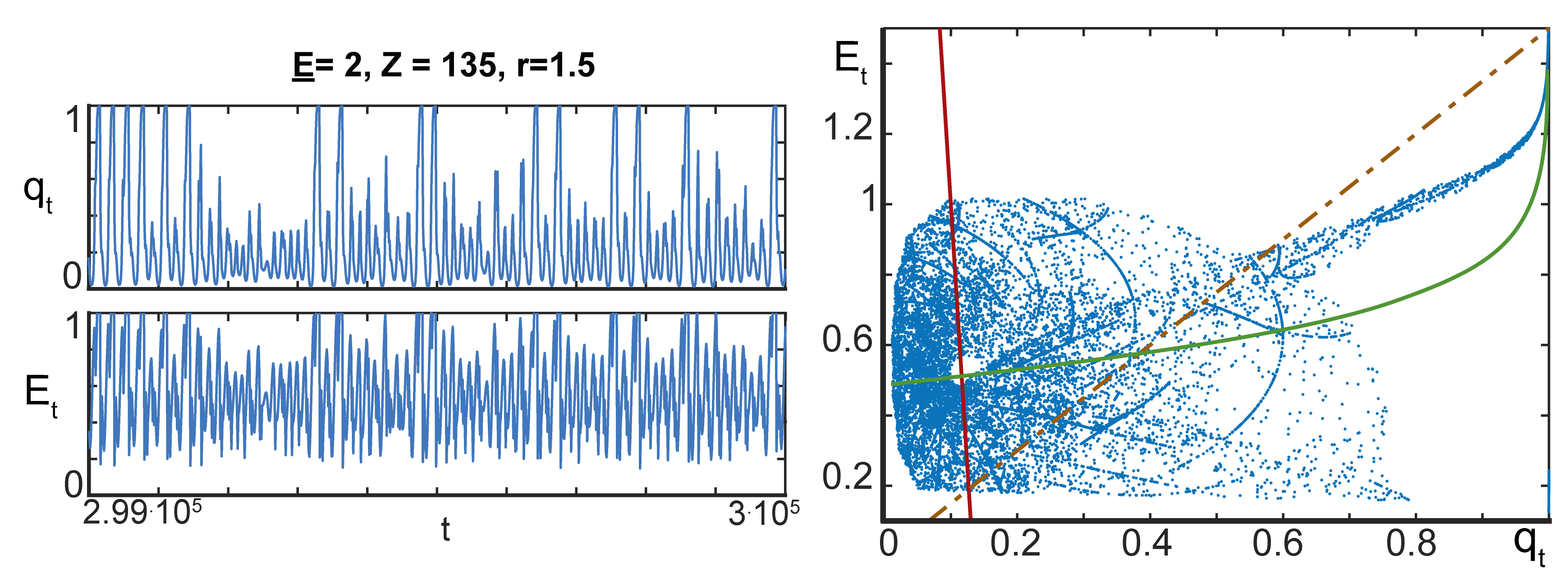}
\caption{Complex behaviour of orbits generated by (\ref{eq:mean_field_adaptive}). {\it Top panels} correspond to  $z=210$, $r=1$, and {\it bottom panels} show dynamics of (\ref{eq:mean_field_adaptive}) for  $z=135$, $r=1.5$. Other parameters of the model are: $w=1.5$, $p=0.1$, $\overline{E}=4$, $\underline{E}=2$, $\varepsilon=0.05$. Green curves show (\ref{eq:sigma_inv}), solid red line shows $E^\ast=\overline{E}-rq^\ast/\varepsilon$, and dashed orange line corresponds to $E^\ast=rq^\ast$. Intersection of the solid green and red curves above the dashed curve reveals equilibrium (\ref{eq:fixed_point_2}). {\it Left panels} present the evolution of $q_t$ and $E_t$ as functions of $t$. {\it Right panels} show the values of pairs $(q_t,E_t)$ for $t\in[2.8\cdot 10^5,3\cdot 10^5]$.}\label{fig:complicated_intermittent}
\end{figure}

All orbits of model (\ref{eq:1.41}) with $q_0\neq 0$, as ensured by Proposition \ref{prop:no_spikes}, converge monotonically to a single non-zero steady state regardless of how large the values of $z$ become. The spectrum of orbits in model (\ref{eq:mean_field_adaptive}) is different. Our numerical experiments demonstrated that, in addition  equilibria, the model is capable of generating periodic orbits too. Moreover, for a broad range of parameters it produced  complicated and apparently chaotic motions.  Examples of these complicated motions are shown in Fig. \ref{fig:complicated_intermittent}.
Observe that the model parameters corresponding to the trajectories in Fig. \ref{fig:complicated_intermittent} satisfy statement 2) of Proposition \ref{prop:energy_model}. In this case, at most two equilibria may exist. As we can see from Fig. \ref{fig:complicated_intermittent}, these equilibria (fixed points (\ref{eq:fixed_point_1}) and (\ref{eq:fixed_point_2})) are not attracting the orbits, and trajectories appear to be chaotic with some apparent intermittency.

\begin{figure}
\centering
\includegraphics[width=\textwidth]{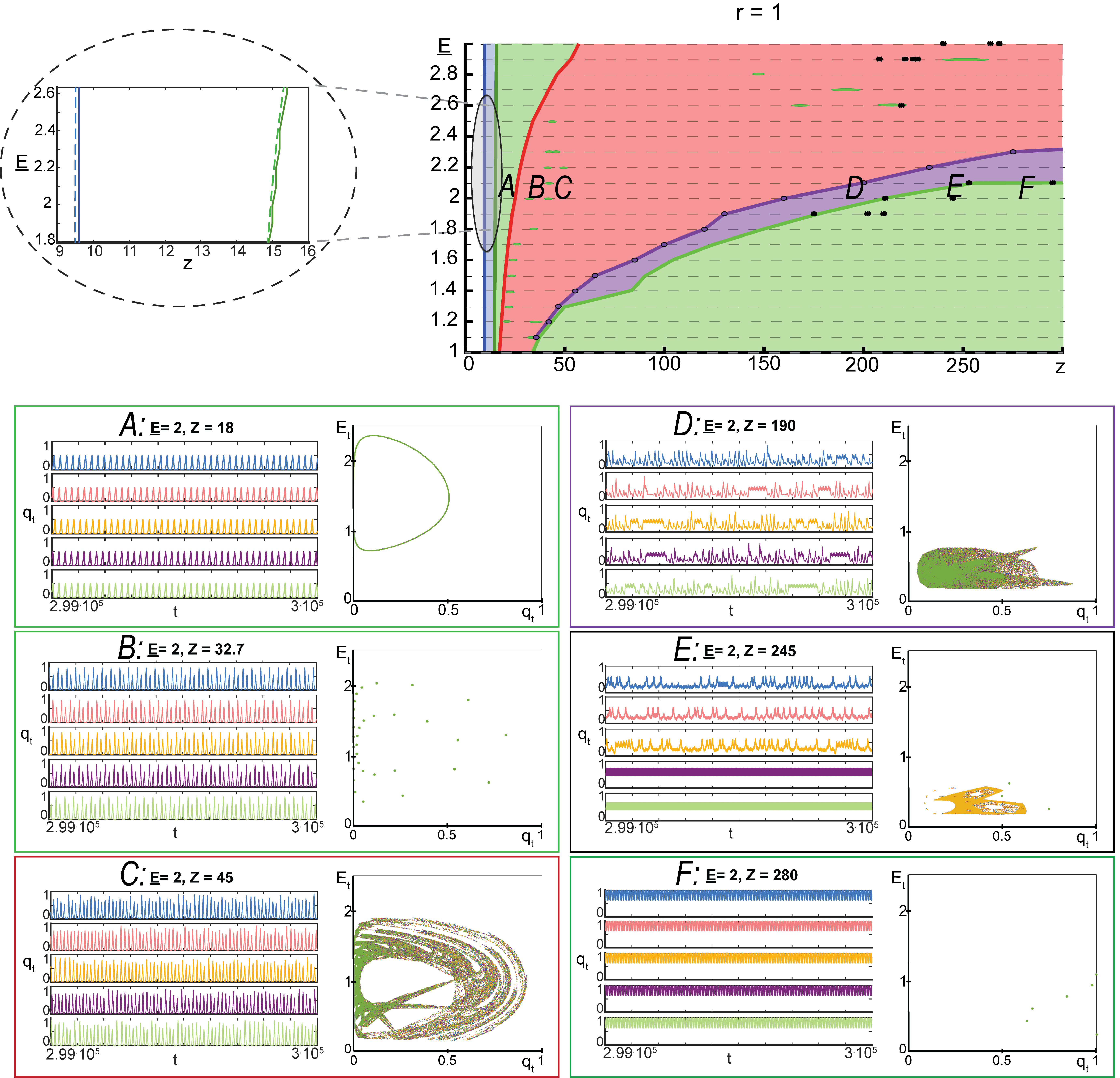}
\vspace{2mm}
\caption{Complex dynamics of model (\ref{eq:mean_field_adaptive}). {\it Top row}:  parametric portrait of the qualitative dynamics of model (\ref{eq:mean_field_adaptive}) in  the domain $\{(\underline{E},z) | \underline{E}\in[1,3], \ z\in (0, 300] \}$ at $r=1$, $w=1.5$, $p=0.1$, $\overline{E}=4$, $\varepsilon=0.05$. White areas show parameter regions in which only one fixed point,  (\ref{eq:fixed_point_1}), was detected. This fixed point is an attractor. Blue region bordering the white one corresponds to the case in which fixed point (\ref{eq:fixed_point_1}) becomes a repeller and  second equilibrium, (\ref{eq:fixed_point_2}), emerges. Equilibrium (\ref{eq:fixed_point_2}) is locally asymptotically stable. Green areas are the domains in which an attracting periodic orbit was detected. Red and violet domains correspond to regions where complex chaotic-like dynamics was observed. Black stars, $\ast$, indicate observed co-existence of multiple attractors. Inlet linked to the gray area shows theoretical estimates of transition boundaries (dashed blue and green lines) relative to the ones observed in experiments. {\it A,B,C,D,E,F}: typical dynamics observed in the corresponding parametric regions.}\label{fig:R1}
\end{figure}

\begin{figure}
\centering
\includegraphics[width=\textwidth]{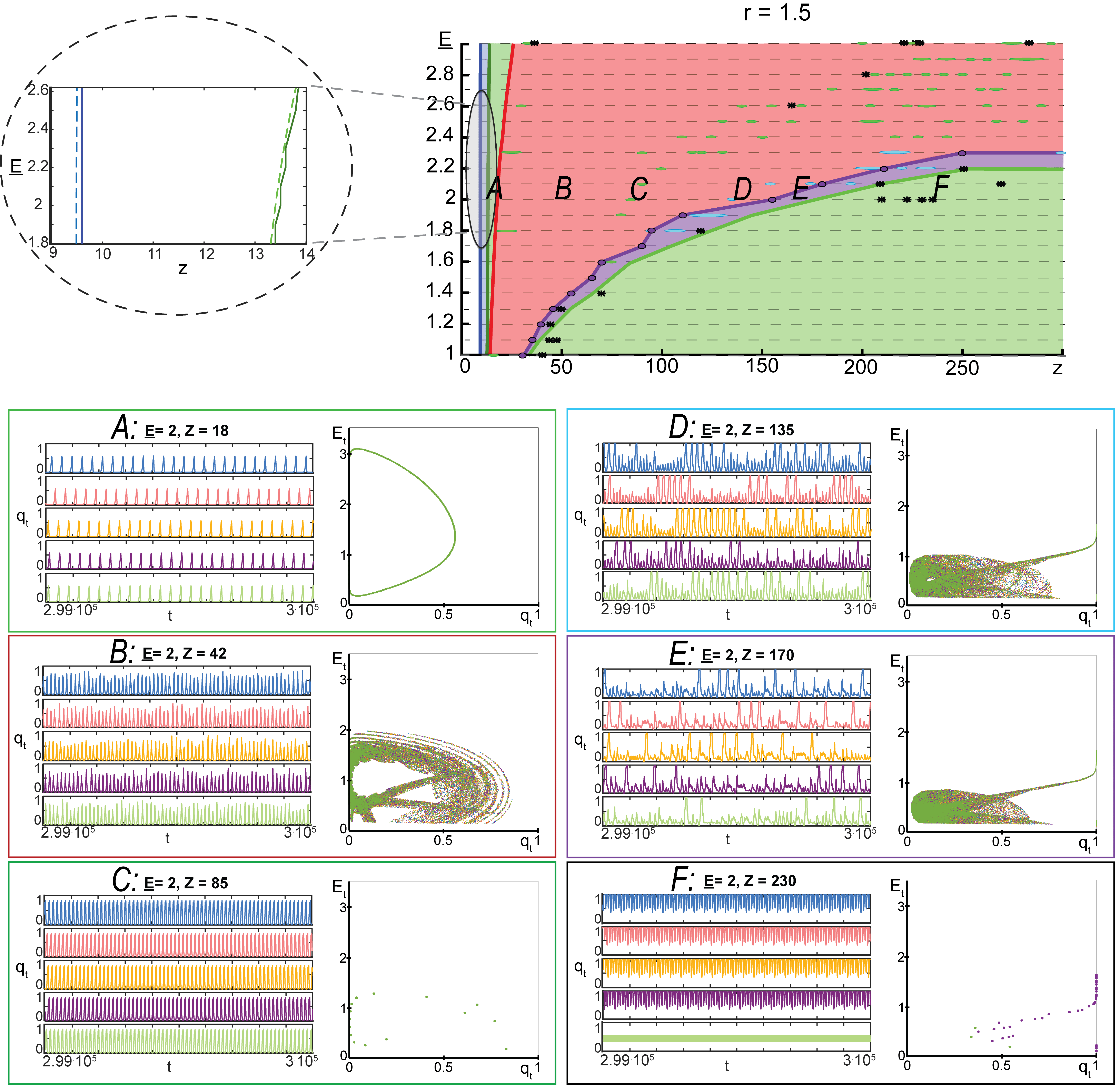}
\vspace{2mm}
\caption{Complex dynamics of model (\ref{eq:mean_field_adaptive}). {\it Top row}:  parametric portrait of the qualitative dynamics of model (\ref{eq:mean_field_adaptive}) in  the domain $\{(\underline{E},z) | \underline{E}\in[1,3], \ z\in (0, 300] \}$ at $r=1.5$, $w=1.5$, $p=0.1$, $\overline{E}=4$, $\varepsilon=0.05$. White areas show parameter regions in which only one fixed point,  (\ref{eq:fixed_point_1}), was detected. This fixed point is an attractor. Blue region bordering the white one corresponds to the case in which fixed point (\ref{eq:fixed_point_1}) becomes a repeller and  second equilibrium, (\ref{eq:fixed_point_2}), emerges. Equilibrium (\ref{eq:fixed_point_2}) is locally asymptotically stable. Green areas are the domains in which an attracting periodic orbit was detected. Red and violet domains correspond to regions where complex chaotic-like dynamics was observed. Black stars, $\ast$, indicate observed co-existence of multiple attractors. Turquoise blue islands mark regions in which burst-like trajectories were observed. Inlet linked to the gray area shows theoretical estimates of transition boundaries (dashed blue and green lines) relative to the ones observed in experiments. {\it A,B,C,D,E,F}: typical dynamics observed in the corresponding parametric regions.}\label{fig:R1_5}
\end{figure}

\begin{figure}
\centering
\includegraphics[width=\textwidth]{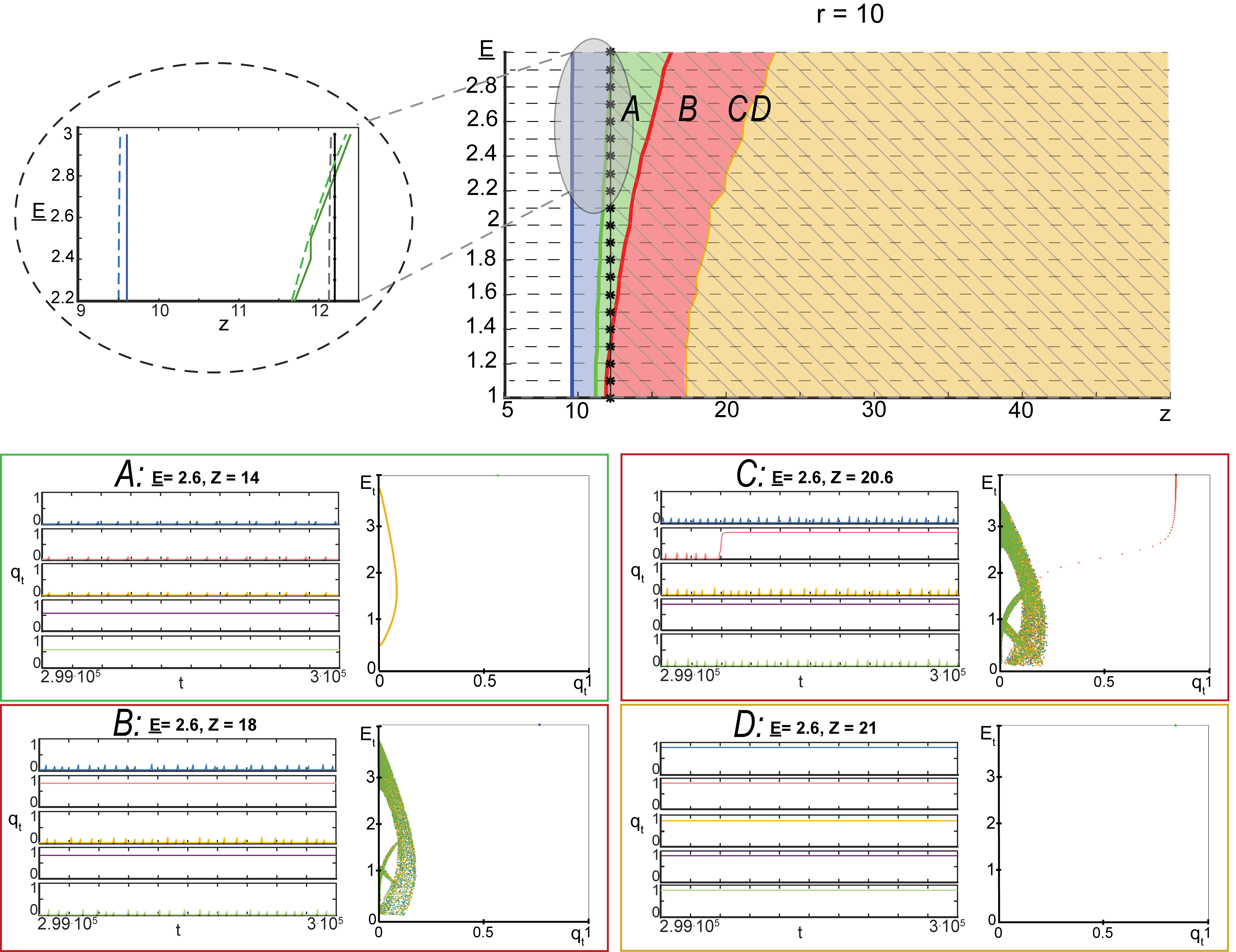}
\vspace{2mm}
\caption{Complex dynamics of model (\ref{eq:mean_field_adaptive}). {\it Top row}:  parametric portrait of the qualitative dynamics of model (\ref{eq:mean_field_adaptive}) in  the domain $\{(\underline{E},z) | \underline{E}\in[1,3], \ z\in (0, 50] \}$ at $r=10$, $w=1.5$, $p=0.1$, $\overline{E}=4$, $\varepsilon=0.05$. White areas show parameter regions in which only one fixed point,  (\ref{eq:fixed_point_1}), was detected. This fixed point is an attractor. Blue region bordering the white one corresponds to the case in which fixed point (\ref{eq:fixed_point_1}) becomes a repeller and  second equilibrium, (\ref{eq:fixed_point_2}), emerges. Equilibrium (\ref{eq:fixed_point_2}) is locally asymptotically stable. Green areas are the domains in which an attracting periodic orbit was detected. Red area corresponds to regions where complex chaotic-like dynamics was observed. Black line indicates a boundary beyond which co-existence of multiple attractors was observed consistently in all experiments.  Inlet linked to the gray area shows estimates of transition boundaries (dashed blue ($\overline{z}_2$ in (\ref{eq:critical})),   green, and black ($\overline{z}_3$ in (\ref{eq:critical})) lines) relative to the ones observed in experiments.  Yellow region shows domain in which all trajectories converged to (\ref{eq:fixed_point_3}). {\it A,B,C,D}: typical dynamics observed in the corresponding parametric regions.}\label{fig:R10}
\end{figure}

In order to gain additional insight into the model's dynamics, we numerically explored asymptotic regimes of (\ref{eq:mean_field_adaptive}) for varying values of $z$, $\underline{E}$, and $r$. Other parameter were as follows: $\varepsilon=0.05$, $w=1.5$, $\overline{E}=4$, $p=0.1$. Outcomes of these experiments are summarized in Figs. \ref{fig:R1} -- \ref{fig:R10} (see Materials and Methods for details of the steps taken to produce these figures). In these experiments, the values of $\underline{E}$ were chosen from a uniform equispaced grid of $21$ points in the interval $[1,3]$. This grid is shown as grey dashed horizontal lines in Figs. \ref{fig:R1} -- \ref{fig:R10}. Parameter $z$ was varying adaptively (increments ranged from $0.1$ in the intervals $(0,50]$ and $(200,300]$  to $5$ in the interval $(50,200]$). For these values of model parameters, we assessed the type of the model's asymptotic dynamics and mapped these onto relevant parametric regions. These regions are shown with different colour in Figs. \ref{fig:R1} -- \ref{fig:R10}.

According to Figs. \ref{fig:R1}--\ref{fig:R10}, development and evolution of the dynamics of (\ref{eq:mean_field_adaptive}) follows a robust pattern. For a fixed value of $\underline{E}\in[1,3]$ and $z$ small, trajectories of the model converge to a unique attractor (fixed point \ref{eq:fixed_point_1}). This attractor corresponds to the system's state in which no elements/neurons are excited. When the value of $z$ increases and exceeds $\overline{z}_2$ (specified in (\ref{eq:critical}) and shown as blue dashed lines in \ref{fig:R1}--\ref{fig:R10}), equilibrium (\ref{eq:fixed_point_1}) becomes a repeller and a second equilibrium emerges (fixed point (\ref{eq:fixed_point_2}). Numerical evaluation of the eigenvalues of the Jacobian at this equilibrium showed that it is locally asymptotically stable. Further increases of $z$ lead to that fixed point (\ref{eq:fixed_point_2}) loses stability thorough the Neimark-Sacker bifurcation, and an attracting periodic orbit emerges. The boundary of this transition is depicted as green dashed lines in Figs. \ref{fig:R1}--\ref{fig:R10}. As we keep increasing the values of $z$, non-trivial and complex dynamics eventually occur (red area in Figs. \ref{fig:R1}--\ref{fig:R10}). Complex orbits and behavior persist over intervals of values of $z$. Notice that  some of the complex trajectories shown in panel $C$, \ref{fig:R10} eventually converge to the stable equilibrium specified by (\ref{eq:fixed_point_3}). This is an empirical manifestation of slow relaxations and critical delays in model (\ref{eq:mean_field_adaptive}) \cite{Gorban:1981}, \cite{Gorban:2004}.  For $z$ sufficiently large, these complex orbits disappear and reduce to periodic orbits, Figs. \ref{fig:R1}, \ref{fig:R1_5}, or mere equilibria, Fig. \ref{fig:R10}.

Mean-field bursting dynamics, shown e.g. in panels D and E in Figs. \ref{fig:R1}, \ref{fig:R1_5}, resembles that of the population bursts observed in live neuronal evolving cultures. An important factor in successful replication of this behavior was the energy variable, $E_t$, coupled with the energy-dependent activation probability $p\sigma(E_t,\underline{E})$. The mean-field model, however, does not capture spatial effects and as such is only a rough approximation of activity propagation in neuronal cultures. In the next section we extend the proposed mean-filed model (\ref{eq:mean_field_adaptive}) to a multi-agent network with randomized energy-dependent activation and numerically assess relevant parameters of its dynamics, including distributions of sizes and durations of firing avalanches.

\paragraph{Multi-agent model of neuronal excitation}\label{sec:aggregated}

As a natural extension of (\ref{eq:mean_field_adaptive}) we considered a connected network of $N$ neurons of which the activity is governed by exogenous energy variables. The network's topology is defined by its adjacency matrix, $C$, whose elements $c_{ij}$ are:
\[
c_{ij}=\left\{\begin{array}{ll}1, & \mbox{if there is a link from the} \ i\mbox{-th} \ \mbox{node to the} \ j\mbox{-th}\\
                               0, & \mbox{otherwise}.  \end{array} \right.
\]
No  links from a node to itself are permitted, but cycles are allowed. For simplicity, all links in the network have been assigned equal weights of which the value was assumed to be $1$. For the given adjacency matrix $C$, we determined the average number of inputs, $\langle{N}_{\mathrm{in}}\rangle$, and the average number of outputs $\langle{N}_{\mathrm{out}}\rangle$
\[
\langle{N}_{\mathrm{in}}\rangle=\frac{1}{N} \sum_{i=1}^N \sum_{j=1}^N c_{ji}, \ \langle{N}_{\mathrm{out}}\rangle=\frac{1}{N} \sum_{i=1}^N \sum_{j=1}^N c_{ij}.
\]

Each $i$-th node in the network is described by two variables: the activity variable, $q_{i,t}\in\Real$, and the energy variable, $E_{i,t}\in\Real$. Dynamics of these variables is defined as follows:
\begin{equation}\label{eq:network_dynamics}
\begin{aligned}
q_{i,t+1} & = a_{i,t}; \\ 
E_{i,t+1} & = (1-\varepsilon) E_{i,t} + \varepsilon \overline{E} - a_{i,t} E^{\mathrm{fire}}_{i,t},\\
\end{aligned}
\end{equation}
where
\[
E^{\mathrm{fire}}_{i,t} =  \frac{r_1 \sum_{j=1}^N  c_{ij}}{\langle{N}_{\mathrm{out}}\rangle} + \frac{r_2 \sum_{j=1}^N  c_{ji} q_{j,t}}{\langle{N}_{\mathrm{in}}\rangle},
\]
and
\[
\begin{aligned}
a_{i,t}&=\left\{\begin{array}{ll}
                 1 \mbox{ with probability } \  p^{\mathrm{fire}}_{i,t}, & \mbox{if} \ E_{i,t}> E^{\mathrm{fire}}_{i,t};\\
                 0, & \mbox{otherwise},
                \end{array}\right.\\
p^{\mathrm{fire}}_{i,t}&=(1-p \sigma(E_{i,t},\underline{E}))^{1+\sum_{j=1}^n c_{j i} q_{j,t}}.
\end{aligned}
\]
The variables $q_{i,t}$ take values in the set $\{0,1\}$, and $E_{i,t}$ are in the interval $[0,\overline{E}]$. The function $\sigma(\cdot,\underline{E})$ is as in (\ref{eq:mean_field_adaptive}).

Phenomenological motivation for the dynamics of individual nodes in model (\ref{eq:network_dynamics}) is similar to that of the mean-field model, (\ref{eq:mean_field_adaptive}). There are, however, several key differences. The evolution of variables $q_{i,t}$ and $E_{i,t}$ in (\ref{eq:network_dynamics}) depends explicitly on the network topology and activity of the node's neighboring cells. The energy balance equation, the second equation in (\ref{eq:network_dynamics}), accounts for the costs of transmitting active signals at the neuron's input (term ${r_2 \sum_{j=1}^N  c_{ji} q_{j,t}}{\langle{N}_{\mathrm{in}}\rangle}^{-1}$) and generating activity signals on the neuron's output (term ${r_1 \sum_{j=1}^N  c_{ij}}{\langle{N}_{\mathrm{out}}\rangle}^{-1}$). If the node's energy level is insufficient to trigger a spike, $E_{i,t}\leq E^{\mathrm{fire}}_{i,t}$,  then no spikes will generated. The latter property is difficult to fully capture at the level of the mean-field approximation, as low sub-threshold values of the bulk energy do not necessarily imply absence of activity at the level of individual neurons (cf. Proposition \ref{prop:energy_model}, alternative $3$, and Fig. \ref{fig:critical_points}).

In our numerical experiments we focused on fully connected networks for which $c_{ij}=1-\delta_{ij}$, where $\delta_{ij}$ is the Kronecker's delta. We also observed that adding a fraction of inhibitory connections does not altar the network dynamics qualitatively. These simplifications are consistent with the approaches used in earlier works \cite{Levina:2007}. The model parameters where set as follows:
\[
p=0.01, \ \underline{E}=2, \ \overline{E}=4, \ w=1.5, \ \varepsilon=0.0025, \ N=625,
\]
and parameters $r_1$ and $r_2$ varied in the intervals $[1,1.5]$ and $[4,6]$, respectively.

\begin{figure}[!]
\centering\noindent
\includegraphics[width=\textwidth]{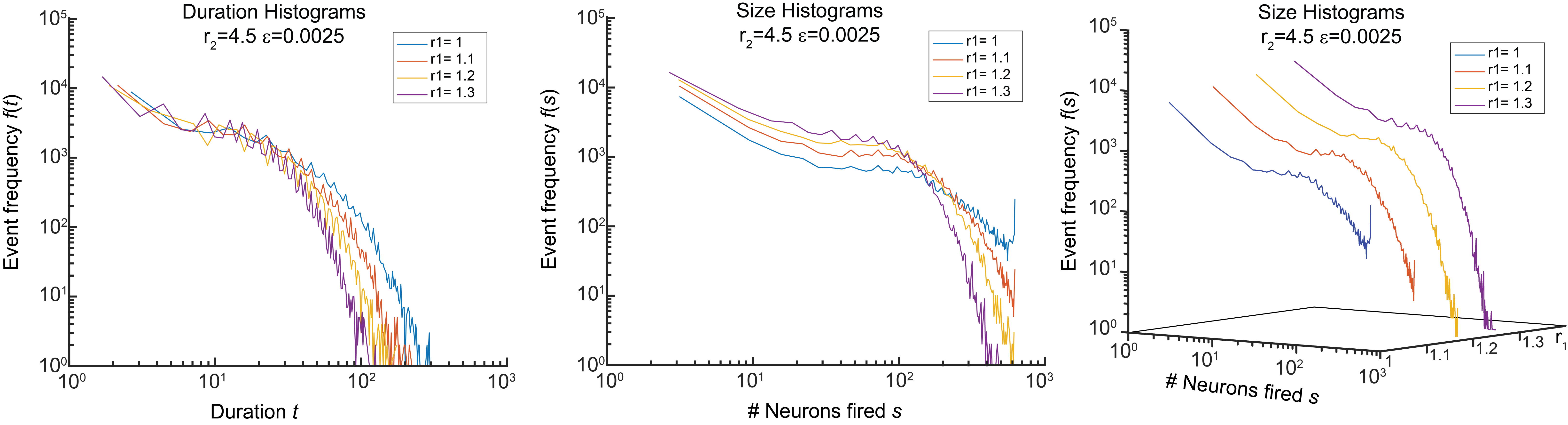}
\includegraphics[width=\textwidth]{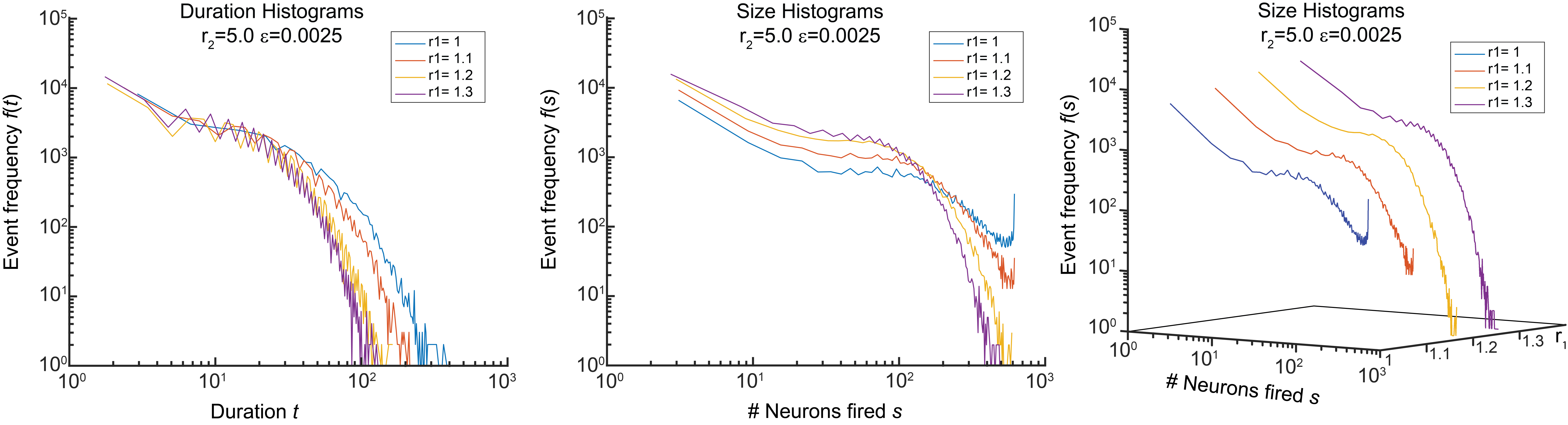}
\includegraphics[width=\textwidth]{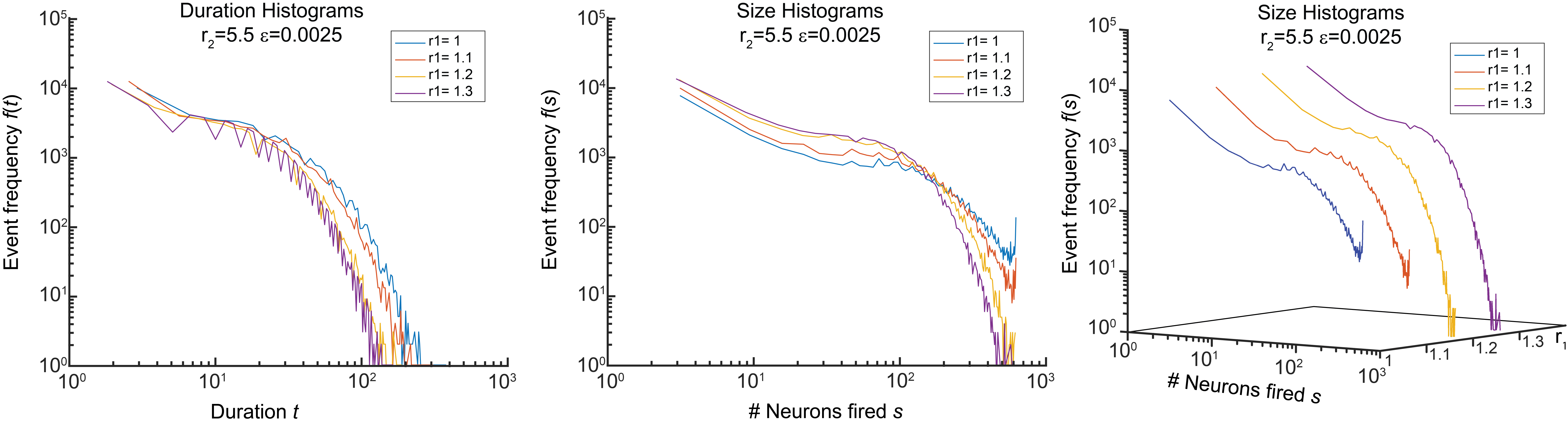}
\includegraphics[width=\textwidth]{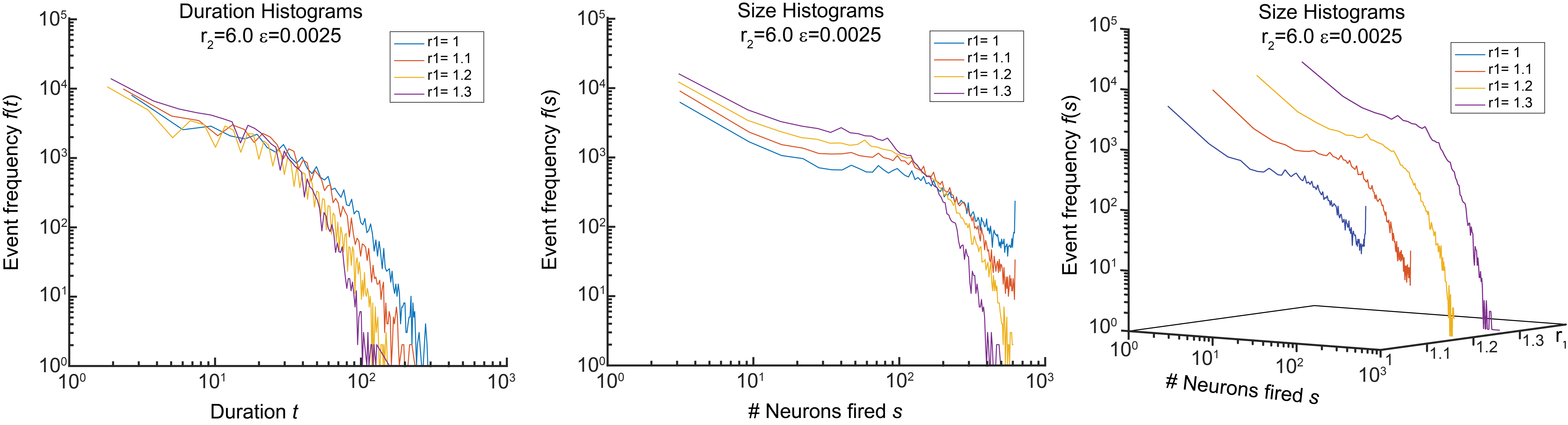}
\caption{Sizes and durations of avalanches for different parameters of the model.}
\label{fig:avalanches_exhaustive}
\end{figure}

\begin{figure}[!]
\centering\noindent
\includegraphics[width=0.5\textwidth]{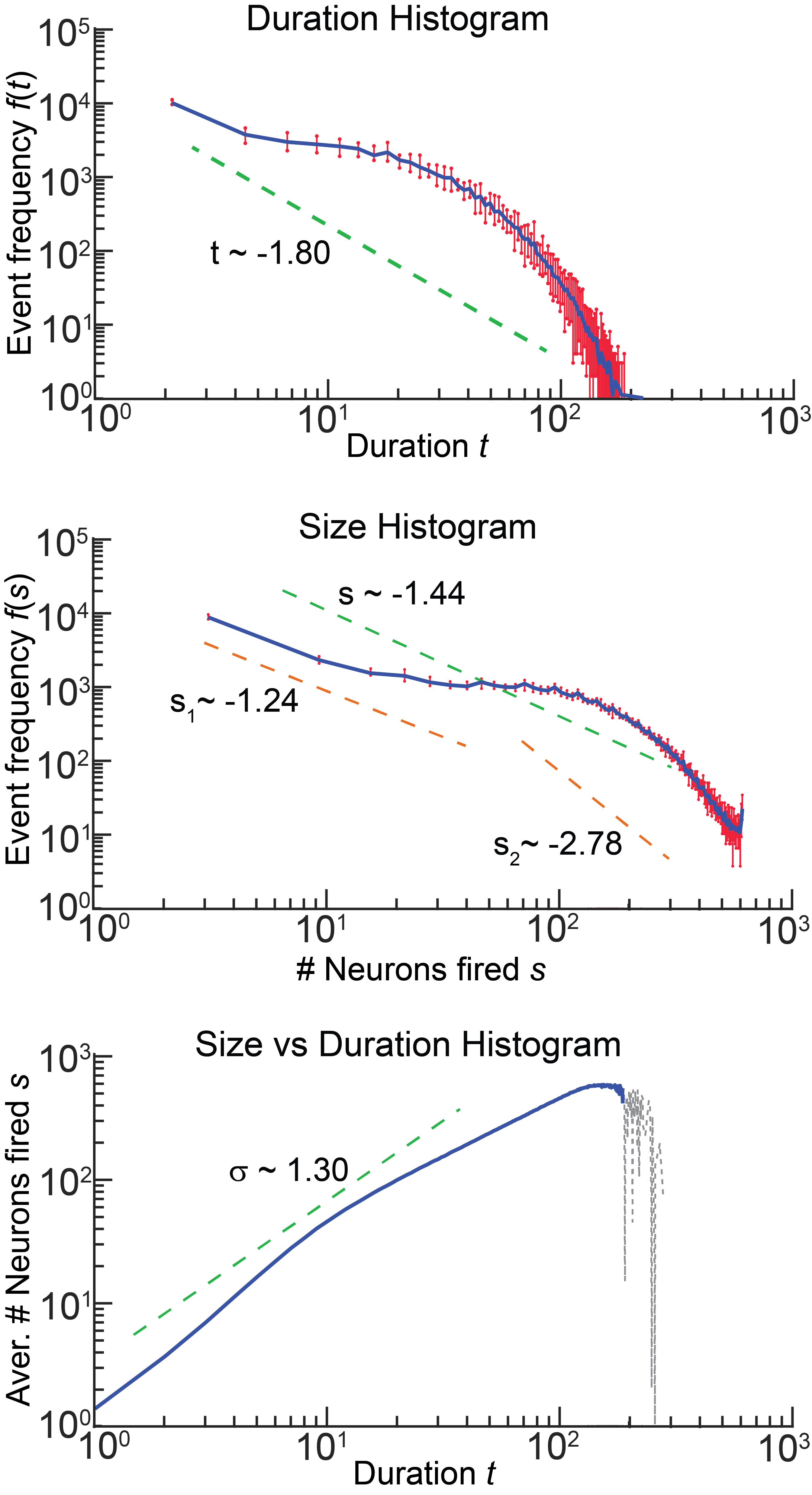}
\caption{Statistics of sizes and durations of avalanches for $r_2=5.5$, $r_1=1.1$, and their estimated exponents. Red bars indicate intervals within which the values of empirical frequencies varied over $10$ different simulations. Blue curves show empirical means. Dashed gray curve in the bottom plot shows a part of the curve corresponding to events for which the data was limited to just few records.}
\label{fig:avalanches_statistics}
\end{figure}

We simulated forward orbits of model (\ref{eq:network_dynamics}) for various initial conditions and parameter values, and determined sizes and durations of avalanches of firing events. In our experiments, the avalanches were defined as events corresponding to the intervals $T_j=[t_{j},t_{j+1}]$ of the network nonzero firing activity such that $\sum_{i=1}^{N}q_{i,t}>0$ for all $t\in T_j$ and  $\sum_{i=1}^{N}q_{i,t_{j-1}}=\sum_{i=1}^{N}q_{i,t_{j+1}+1}=0$. Each orbit was simulated for $10^6$ time steps, with $q_{i,0}=0$, $i=1,\dots,N$ and $E_{i,0}$, $i=1,\dots,N$ chosen randomly in the interval $[0.5 \underline{E}, \overline{E}]$. For each orbit, we gathered statistics of sizes and durations of the observed avalanches.  A brief summary of these experiments is shown  in Fig. \ref{fig:avalanches_exhaustive} and Fig. \ref{fig:avalanches_statistics}.
Fig. \ref{fig:avalanches_exhaustive} presents size and duration histograms as a function of parameters $r_1$ and $r_2$ in (\ref{eq:network_dynamics}). As we can see from this figure, energy feedback has the capacity to inhibit system-size events in networks with identical graph topologies. Parameters of this feedback may affect the values of size and duration exponents. In particular, we observed that increading the value of $\varepsilon$ facilitates occurances of system-size events and eventually pushes the system into the super-critical state.

Fig. \ref{fig:avalanches_statistics} shows statistics and estimated exponents of avalanches for $r_2=5.5$, $r_1=1.1$.
The estimated exponents are close to those reported for live neuronal cultures \cite{beggs:04}, \cite{Chiappalone:08}, \cite{Beggs:2012}. Observe that the size and duration curves have noticeable humps (cf. \cite{Beggs:2012}, Fig. 2), albeit different and less prominent exponents than those reported in \cite{Beggs:2012}.

Activity bursts appear to be synchronized with the peaks of the energy function. This bears some similarity with network models in which connections change according to rate-dependent synaptic plasticity \cite{Markram:00}. Analogous behavior was observed in the dynamics of mean-field approximation (\ref{eq:mean_field_adaptive}), Figs. \ref{fig:R1}--\ref{fig:R10}.  


\subsection*{Comparison with empirical data}

\begin{figure}[!h]
\centering\noindent

Empirical data\\
\includegraphics[width=\textwidth]{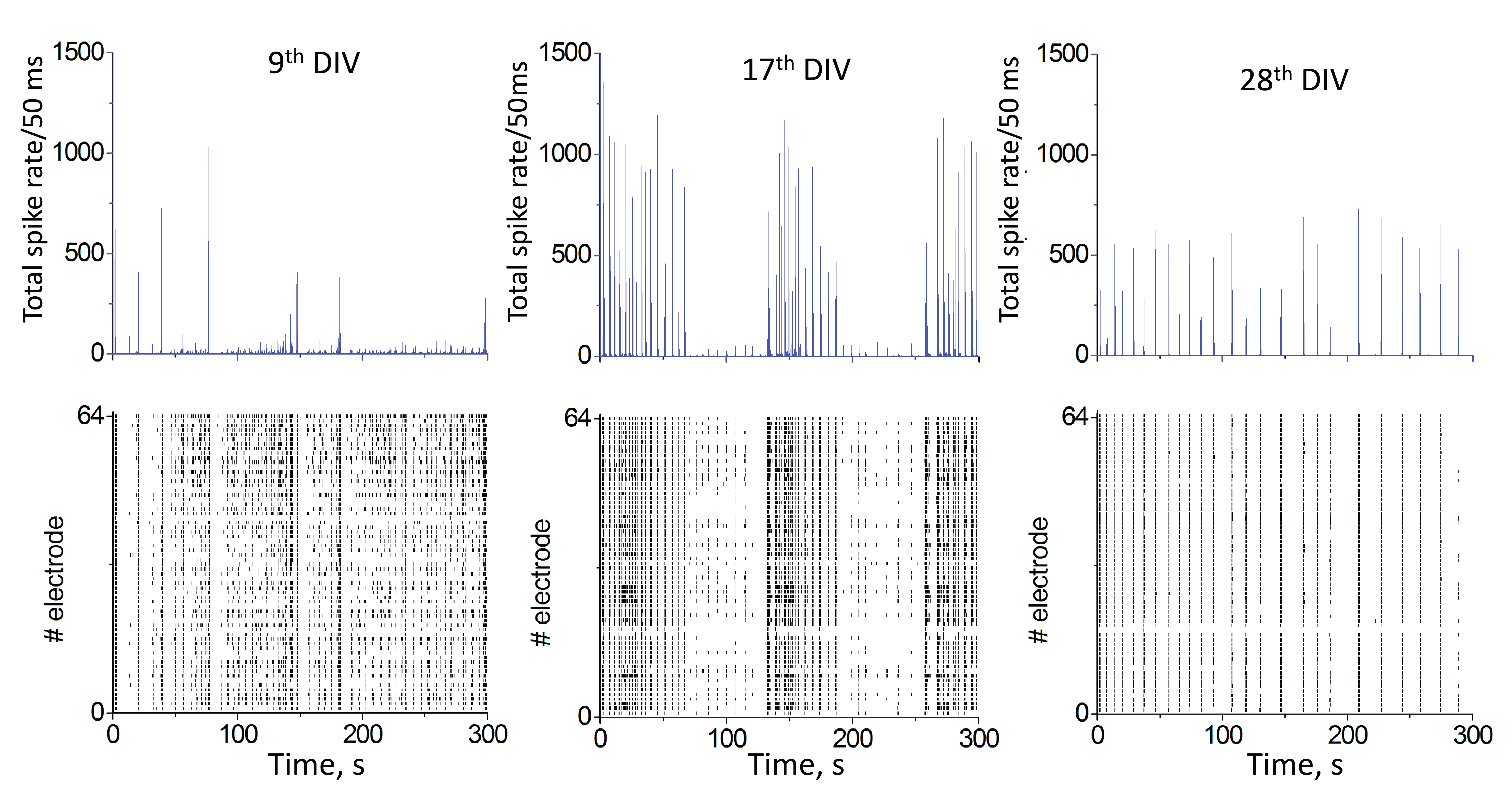}

\vspace{2mm}

Model behavior\\
\vspace{1mm}
\includegraphics[width=.3\textwidth]{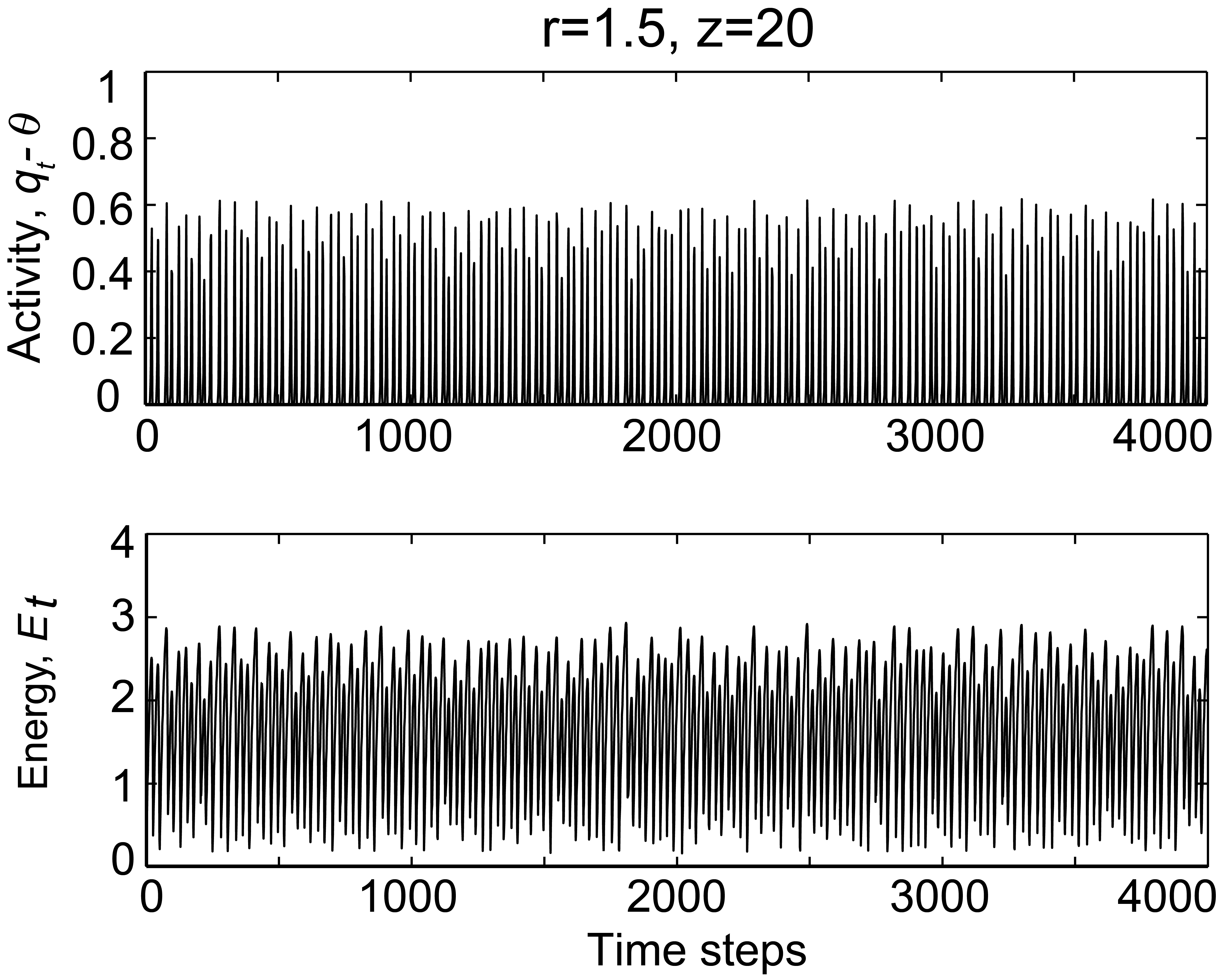}
\includegraphics[width=.3\textwidth]{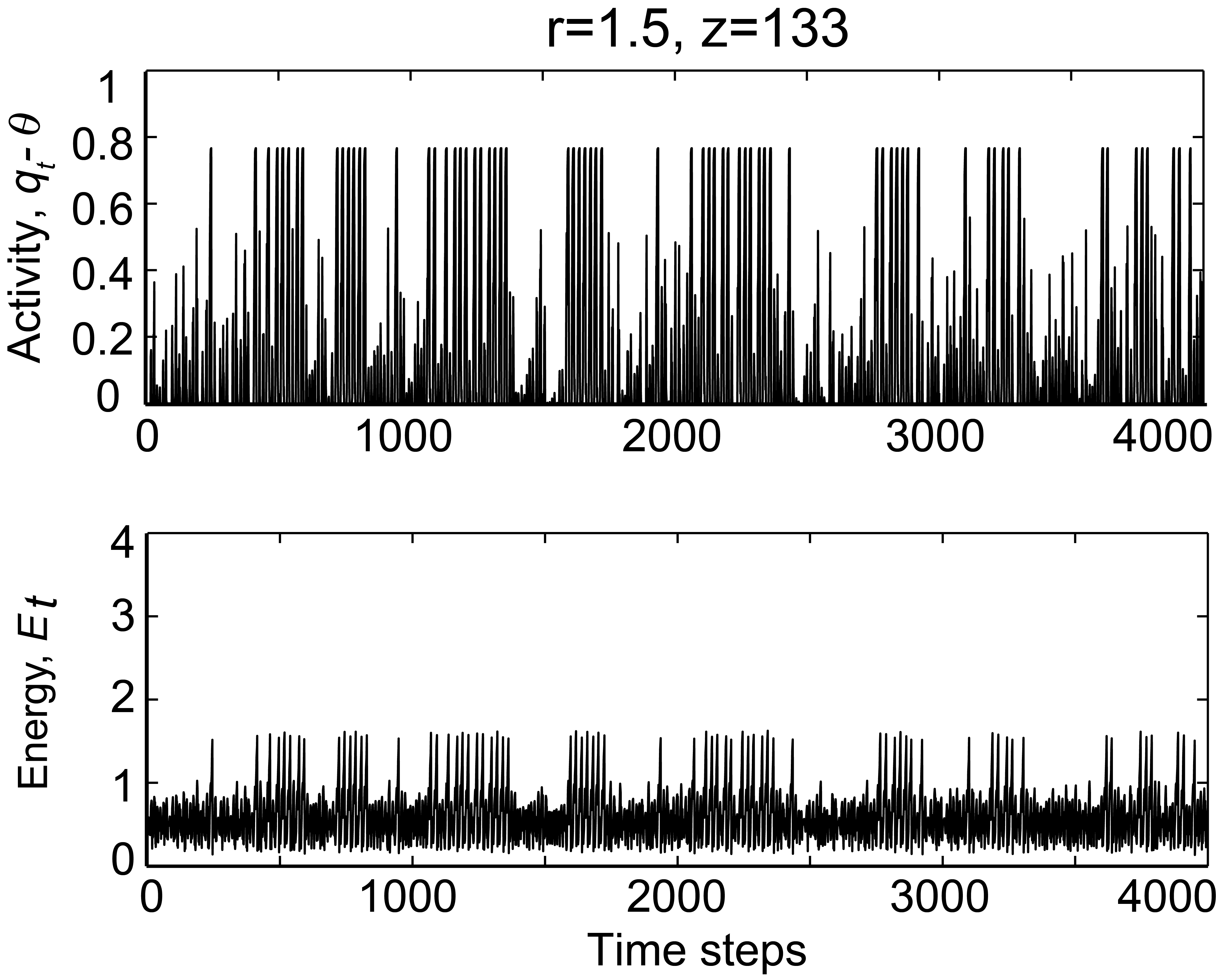}
\includegraphics[width=.3\textwidth]{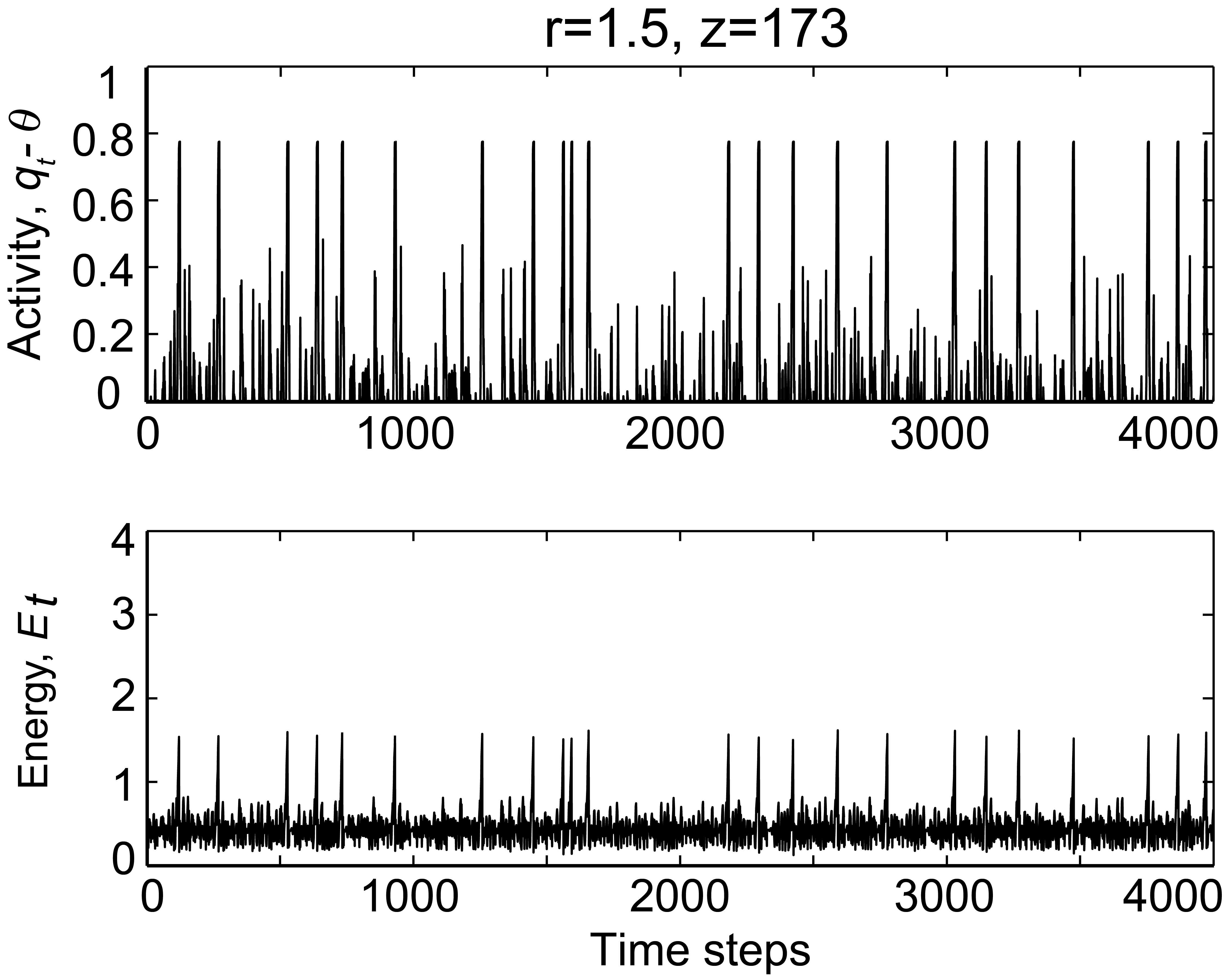}
\caption{{\it Top row}: typical raster diagrams of spontaneous network activity in live neuronal cultures (in vitro). Left plot corresponds to the $9$th day in development (DIV), middle stands for the $17$th DIV, and rightmost represents the $28$th DIV.  {\it Bottom row}: evolution of model variables for various values of network connectivity parameter, $z$. The value of $r$ is set to $1.5$; the values of the rest of the model parameters have not been changed. The activity variable $q_t$ shown was subjected to threshold subtraction.}\label{fig:cultures_developing}
\end{figure}

The model presented above shows a range of behaviors controlled by just few parameters wich may be explicitly linked with relevant physical  quantities. Plots presented in Figs. \ref{fig:R1} -- \ref{fig:R10} show evolution of activity patterns in the network as a function of connectivity parameter $z$. It would therefore be nice to see how these observations relate to patterns observed in actual cultures. Fig. \ref{fig:cultures_developing} shows typical activity of live developing neuronal cultures over time.

We can see from Fig.~\ref{fig:cultures_developing} that, depending on days in development, networks of living cultures exhibit irregular spiking, bursts, and (nearly) periodic spiking. The mean-field dynamics of the activity variable, $q_t$, in our model shows qualitatively similar regimes. Plots shown in Figs.~\ref{fig:R1}--~\ref{fig:R10}, however, have not been subjected to any threshold subtraction. Fig.~\ref{fig:cultures_developing}, bottom row, shows behavior of the model with threshold subtraction for growing values of $z$. The value of threshold, $\theta$, is set to $1/0.6745$ of the median of $q_t$ over the interval of simulation $[0,4000]$. Increases in values of $z$ model development of synaptic connections over time. Evolution of the variable $q_t$ resembles that of the culture activity. Spikes of $q_t$ in the leftmost plot in are narrow and do not correspond to the network spikes. When the value of $z$ is increased their width grows and they start to appear in packs (middle picture); this corresponds to network bursts shown in the first row. When the value of $z$ is increased further the bursts start to break (the rightmost plot), and when $z$ reaches even larger values no high-amplitude spikes are generated in the model (not shown in the figure). The latter
phenomenon was not observed in live cultures. Network connectivity in cultures, however, does not grow indefinitely. Therefore network activity for extreme values of $z$ maye be discarded as that outside of the model validity zone.

\begin{figure}
\centering

Empirical data\\

\vspace{1mm}
\includegraphics[width=\textwidth]{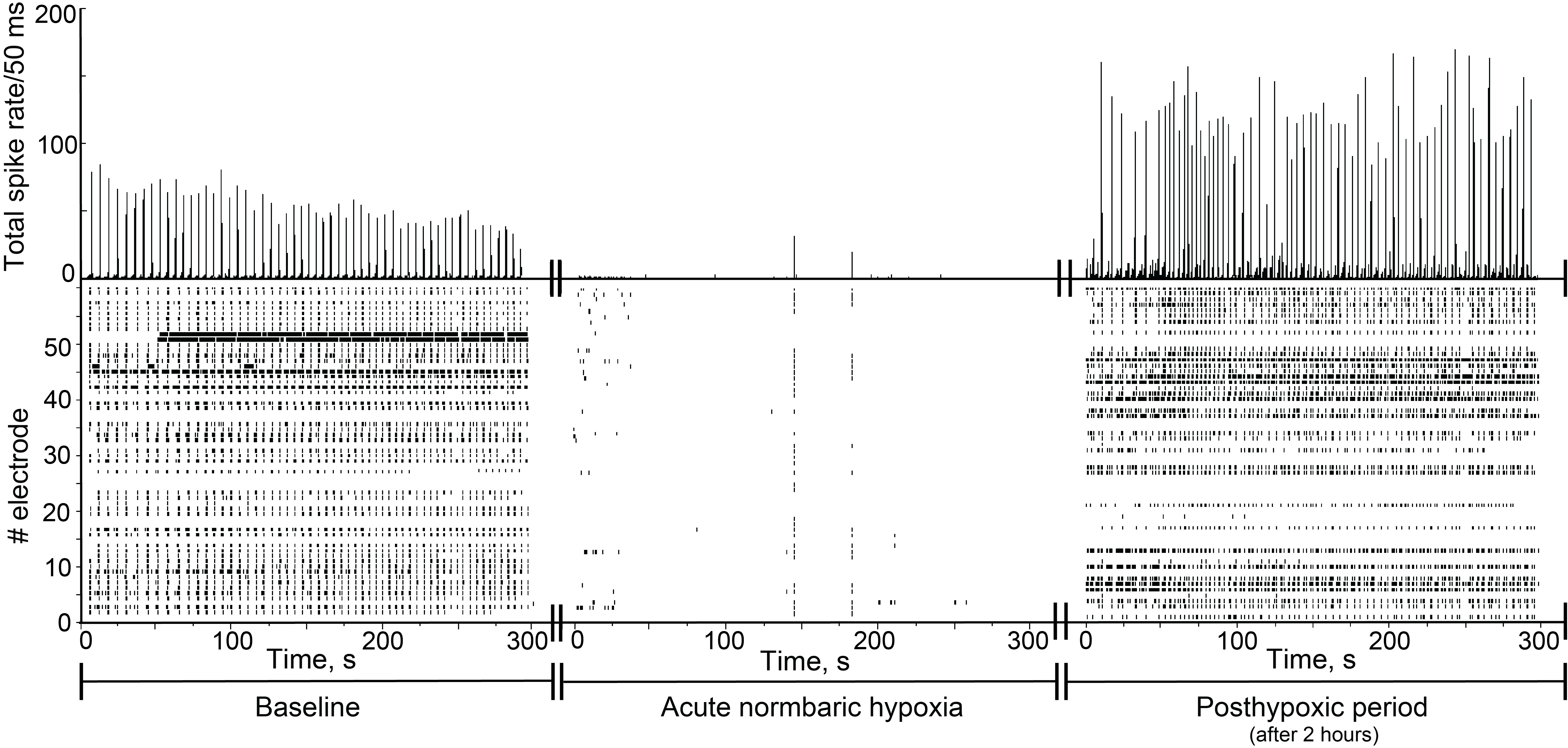}
\vspace{1mm}

Model behavior\\
\vspace{1mm}
\includegraphics[width=0.43\textwidth]{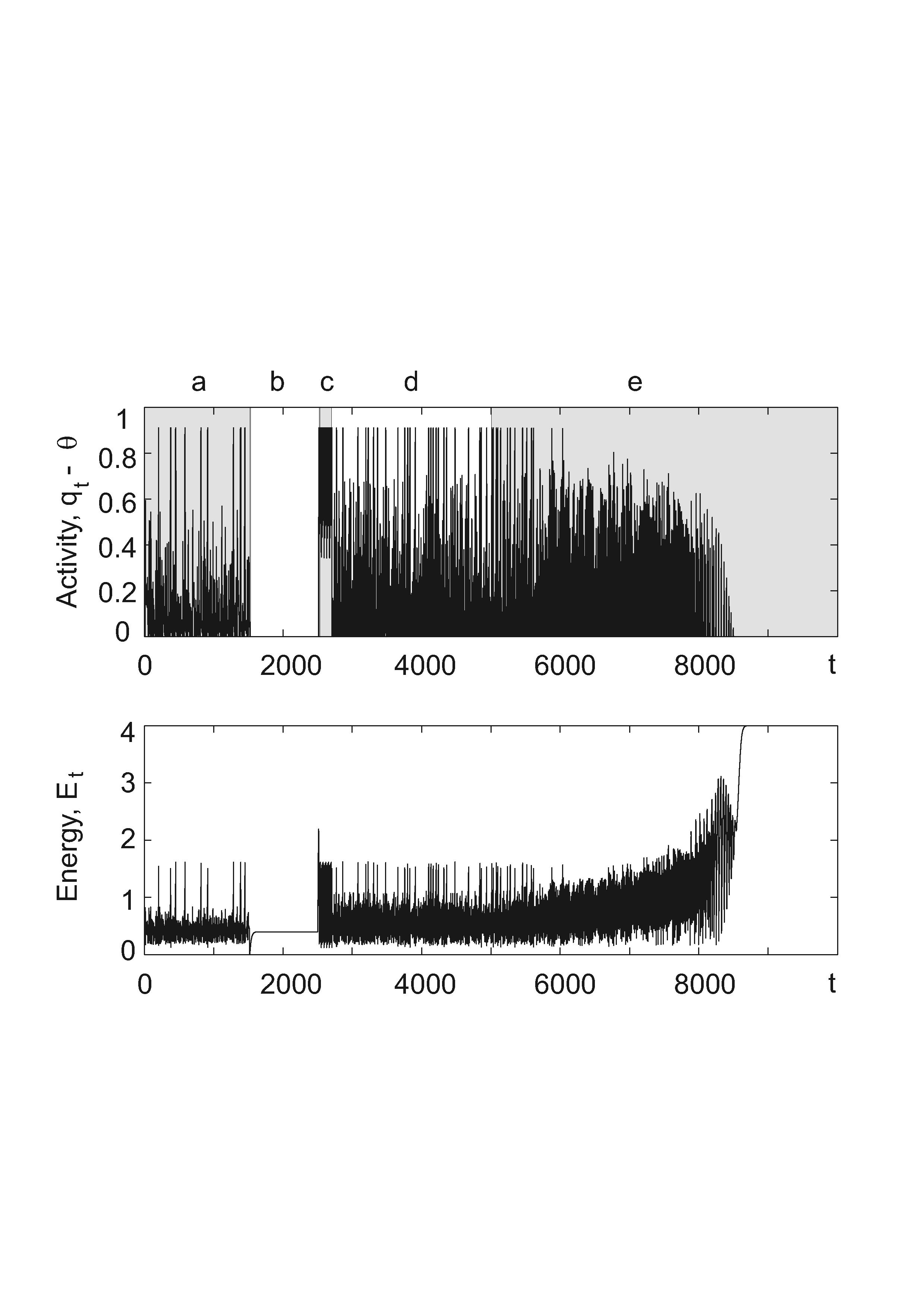} \hspace{5mm}
\includegraphics[width=0.43\textwidth]{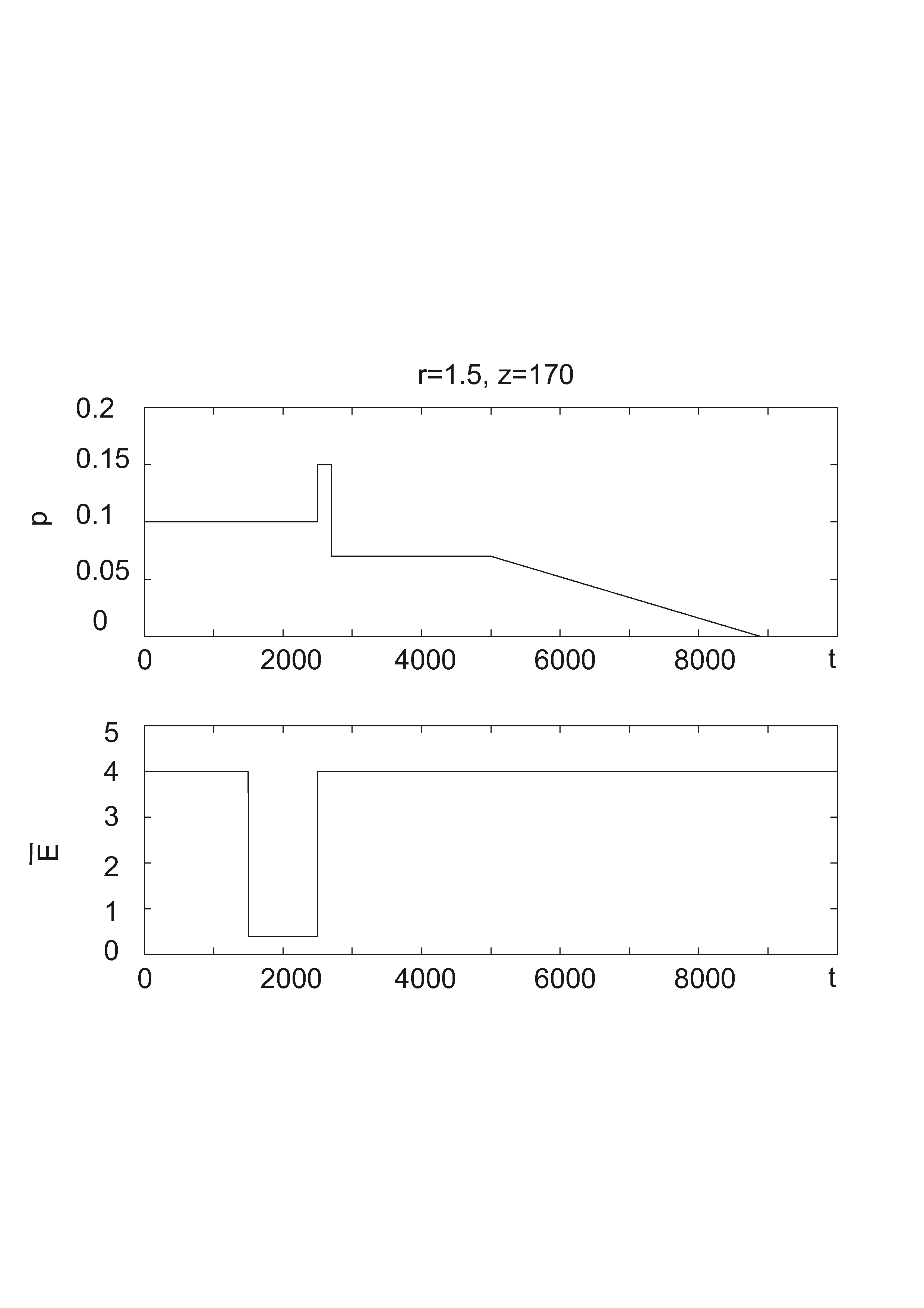}
\vspace{1mm}
\caption{{\it Top panel:} network dynamics before, during, and after acute normobaric hypoxia. {\it Bottom panel}: behavior of the model with phenomenological simulation of the effect of acute oxygen deprivation. Zone $a$ in the left plot corresponds to normal functioning, $b$ shows effect of acute hypoxia, $c$ emulates glutamate induced increase of activity, $d$ and $e$ are the intervals corresponding to glutamate-induced suppression of neurons and degenerative damage.}\label{fig:oxygen}
\end{figure}

Last but not least is that behavior of the model is consistent with the effects of oxygen deprivation in cultures. Fig. \ref{fig:oxygen} shows network activity during and after acute but short oxygen deprivation (see also \cite{Vedunova:14}). In the latter experiments primary cultures of hippocampal cells were subjected to $10$min of acute oxygen deprivation in the $21$st DIV (see Methods). Reoxygenation after short-term hypoxia rapidly restores energy deficit and neuronal ATP levels and increases the release of glutamate. Glutamate is a major excitatory neurotransmitter in mammalian central nervous system. Glutamatergic neurons form the main excitatory system in the brain and play a pivotal role in many neurophysiological functions \cite{Zhouand:2014}. Excessive release of glutamate is homeostatic response to the hypoxia-induced network silence \cite{Meldrum:2000}. On the one hand, it is aimed at restoration of network activity. On the other hand,  it  over-activates  ATP-dependent ion pumps  and changes calcium homeostasis. This in turn leads to a cascade of intracellular events causing neuronal degeneration or excitotoxicity \cite{Dong:2009}, \cite{Nguyen:2011}.

In order to see how the proposed model might capture hypoxia as well as the sequence of  complex biological of changes related to oxygen deprivation the following experiments have been performed. The model, eq (\ref{eq:mean_field_adaptive}), with $r=1.5$, $z=170$, $\varepsilon=0.05$, $w=1.5$, $\underline{E}=2$, $\overline{E}=4$, $p=0.1$ was run for $1500$ steps. Then for the next $1000$ steps the value of $\overline{E}$ was set to $0.4$ and then restored to the nominal level of $4$ for $t>2500$. This corresponds to energy deficit caused by acute oxygen deprivation. At $t=2500$, however, the value of $p$ was increased to $0.15$ to account for glutamate release, and then dropped to the level  $p=0.07$ in the interval $[2700, 5000]$ to emulate glutamate-induced suppression \cite{Mohapatra:2009}. For $t>5000$ the value of $p$ was made to decrease linearly to account for degenerative processes triggered by oxygen deprivation. Model behaviour as well as the evolution of $p$ and $\overline{E}$ over time are shown in Fig. \ref{fig:oxygen}. Overall evolution patterns that the model produces qualtiatively coincide with empirical observation. Notice also that our empirical data showed increase in the level of activity after $2$h of hypoxia. Qualitatively very similar behavior has been reproduced by the model. In this regime, the value of neural excitability parameter $p$ was lower than in the nominal pre-hypoxic state to simulate the combined influence of glutamate-induced suppression. Yet, the network produced larger number of spikes. This can be attributed to network effects and the interplay between activity-energy variables. We also note that other experiments reported e.g. in \cite{Gavello:2012} show that cultures exposed to hypoxia may show reduced activity during oxygen deprivation and after reoxygenation for $2$h (Fig. 5 in \cite{Gavello:2012}). The experimental protocol, however, in \cite{Gavello:2012} differed from ours in that the hypoxia was induced by exposing the cultures to $3\%$ $O_2$ for two hours, and that the age of the cultures was $11$ and $18$ DIV. Variability of development and neurons might also have attributed to differences in observed behavior.

\section*{Materials and Methods}\label{sec:methods}

\subsection*{Cell cultures}

Hippocampal cells were dissociated from embryonic mice 57L/6J (on embryonic day $18$) and plated with a high initial density of approximately $9000$ cells/mm$^2$ on microelectrode arrays (Alpha MED Science, Japan) pre-treated with the adhesion promoting molecule polyethyleneimine (Sigma P3143) according the protocol \cite{Vedunova:13}. The cells were cultured under constant conditions of $35.5^\circ$C, $5\%$ CO2 and  $95\%$ air at saturating humidity in a cell culture incubator (MCO-18AIC, Sanyo) in a medium containing Neurobasal medium (Invitrogen 21103-049) with $2\%$ B27 (Invitrogen 17504-044), $1$ mM L-glutamine (Invitrogen 25030-024) and $0.4\%$ fetal calf serum (PanEco 055) without any antibiotics or antimycotics. Glial growth was not suppressed because glial cells are essential for long-term culture maintenance. One half of the medium was changed every $2$ days.

\subsection*{Electrophysiological methods}

Extracellular potentials were recorded simultaneously through $64$ planar indium tin-oxide (ITO) platinum black electrodes with the integrated MED64 system (Alpha MED Science, Japan). MEAs were $8 \times 8$ $(64)$ with a $50$ $\mu$m $\times$ $50$ $\mu$m electrode size and a $150 \mu m$ spacing, and the sampling rate was $20$ kHz/channel. All of the signal analyses and statistics were performed using custom made software (Matlab$\circledR$).

\subsection*{Spike detection}

Detection of recorded extracellular spikes was based on threshold calculation using the signal median: $T=N_s\sigma$, $\sigma={\mathrm{median}}\left(|x|/0.6745\right)$, where $x$ is the bandpass-filtered ($0.3-8$ KHz) data signal, $\sigma$ is an estimate of the standard deviation of the signal with zero spikes, and $N_s$ is a spike detection coefficient determining the detection threshold \cite{Pimashkin:11}. The standard deviation of Gaussian noise is equal to the median of the absolute values of the signal divided by $0.6745$. $N_s=4$ was used for all data, resulting in a reliable detection of spikes with amplitudes greater than $20$ $\mu$V. The minimal interspike interval was set to $1$ ms. Detected spikes were plotted in raster diagrams.

To detect small network bursts, we calculated the total spiking rate (TSR) accounting for the total number of spikes from all electrodes within 50 ms time bins. A fast appearance of a large number of spikes over all of the electrodes in a small ($50$ ms) time bin was used as a criterion for small burst appearance (see \cite{Pimashkin:11} for more details).

Superbursts in the electrical activity recorded from the multielectrode arrays were detected as follows. First, we defined a Gaussian function with an effective width equal to $10$ s. Next, we iteratively moved that function from the beginning of the recording to the end in $10$ ms time steps and calculated a cross-correlation of the function with the TSR. The resulting cross-correlation indicated how much of the synchronized activity (bursts) was recorded in the $10$ s window. The procedure was identical to the one reported in \cite{Vedunova:13}. To detect superbursts in the spiking activity, we applied threshold detection, in which the threshold was estimated as the spiking superburst detection accuracy coefficient multiplied by the standard deviation of the calculated cross-correlation. The superburst detection accuracy coefficient was found empirically and was equal to $0.4$. All time points that crossed the threshold were defined as the beginnings and endings of the superbursts.

\subsection*{Construction of Figures \ref{fig:R1} -- \ref{fig:R10}}

To construct the figures, we run $3\cdot 10^5$ iterations of model (\ref{eq:mean_field_adaptive}) from $5$ random initial conditions for each relevant set of parameters $\underline{E}$, $z$, $r$, whilst keeping the values of $w,p,\overline{E}$, and $\varepsilon$ constant ($w=1.5$, $p=0.1$, $\overline{E}=4$, $\varepsilon=0.05$). The values of parameter $\underline{E}$ where chosen in the equispaced grid of $21$ points in the interval $[1,3]$. The values of $z$ were  varying adaptively. In the intervals $(0,50]$ and $(200,300]$  these values were taken from  equispaced grids with distances between the grid's nodes being equal to $0.1$. In the interval $(50,200]$ these distances were set to $5$. For each set of model parameters and each initial condition, last $2 \cdot 10^4$ points in each run have been recorded. Points corresponding to orbits from different initial conditions have been collated together (color-coded), plotted in $(q_t, E_t)$ space and stored as .gif files at  \cite{Tanya:GitHub}. This resulted in circa $32 000$ images for $r=1$ and $r=1.5$, and circa $10000$ images for $r=10$.  The resulting figures were visually inspected and classified  into orbits converging to a) single equilibrium (\ref{eq:fixed_point_1}), b) single equilibrium (\ref{eq:fixed_point_2}) c) single periodic orbit, d) complicated sets like the ones shown in Fig. \ref{fig:complicated_intermittent}, and e) multiple attractors. These were color-coded and mapped onto the relevant parametric domains.

\subsection*{Hypoxia modelling}


Hypoxia modelling was conducted by replacing the normoxic culture medium with a medium containing low oxygen (0.37mL/L) for 10 min on day $21$ of culture development in vitro according to the previously described protocol \cite{Vedunova:15}.

\section*{Conclusion}\label{sec:conclusion}

In summary, we have proposed a simple network model explaining burst generation in living culture networks. A distinct feature of our model is presence of a dynamic exogenous energy variable and neuronal activation probability that is made dependent on the energy, like in general models of physiological adaptation \cite{Gorban:15}. We showed that introduction of these modifications already enables to explain evolution of cultures from resting state to population bursts, at least in the mean-field approximation. In accordance to the model, emergence of bursts and spikes is regulated by just few parameters that correspond to network connectivity and efficacy of synaptic transmission. We also note that our energy-based model is complementary to more traditional connectivity-focused approaches \cite{Levina:2007}.

In this particular study, when comparing empirical data with model behavior, the number of days in development has been related to network connectivity. We note, however, that the latter in a broader biological context can depend on various external factors such as e.g. stress \cite{Censi:2011}. The proposed model hence might be able to predict qualitatively the effect of stress and adaptation to stress in neuronal activity.

Large-scale multi-agent simulations demonstrated that these additional variables  are capable of governing  the network's dynamical state and maintaining it at the edge of percolation transition, depending on parameters. The feedback acts as as a mechanism for controlling and maintaining metabolic homeostasis; this enables communication between nodes across the whole network and at the same time prevents network's overload due to excessive propagation of the activity. The energy feedback explicitly suppresses excitation in individual neurons (locally) by disabling high frequency local spike generation. In the other words, it can be viewed as a realization of frequency dependent synaptic plasticity \cite{Markram:00}.

Despite the model qualitatively explains certain phenomena observed in neuronal cultures, the model is simplistic. It does not account for varying strengths of synaptic efficacy as well as for various plasticity mechanisms and their time scales. Accounting for these is the subject of our future work.






\section*{Ethics statement}

C57Bl/6j pregnant mice (E18) were euthanized via cervical dislocation according the recommendations in the Guide for the Care and Use of Laboratory Animals of the Ministry of Health of the Russian Federation. The protocol was approved by the Committee on the Ethics of Animal Experiments of the Privolzhsky Research Medical University (Protocol Number: 09-02.2016). All efforts were made to minimize suffering.

\section*{Acknowledgments}
The work supported by the Ministry of education and science of Russia (Project No. 14.Y26.31.0022).


\section*{Appendix. Proofs of technical statements}\label{sec:appendix}

\subsection*{Proof of Proposition \ref{prop:no_spikes}}

Notice that $a=(1-p)^{z}\in(0,1)$ for all $p\in(0,1), z>0$. Thus $1-a^{q_t}\in[0,1]$ for all $q_t\in[0,1]$, and forward invariance of $[0,1]$ follows. The right-hand side of (\ref{eq:1.41}) is continuous and strictly monotone with respect to $q_t$ on $(0,1]$, with $q_t=0$ being an equilibrium. Hence all forward orbits of this map, i.e. $q_t,q_{t+1}, q_{t+2},\dots$ are monotone, and map (\ref {eq:1.41}) has only fixed points as attractors. Furthermore, the right-hand side of (\ref{eq:1.41}) is strictly concave, which implies that the number of fixed points is at most two.

\begin{figure}
\centering
\includegraphics[width=0.8\textwidth]{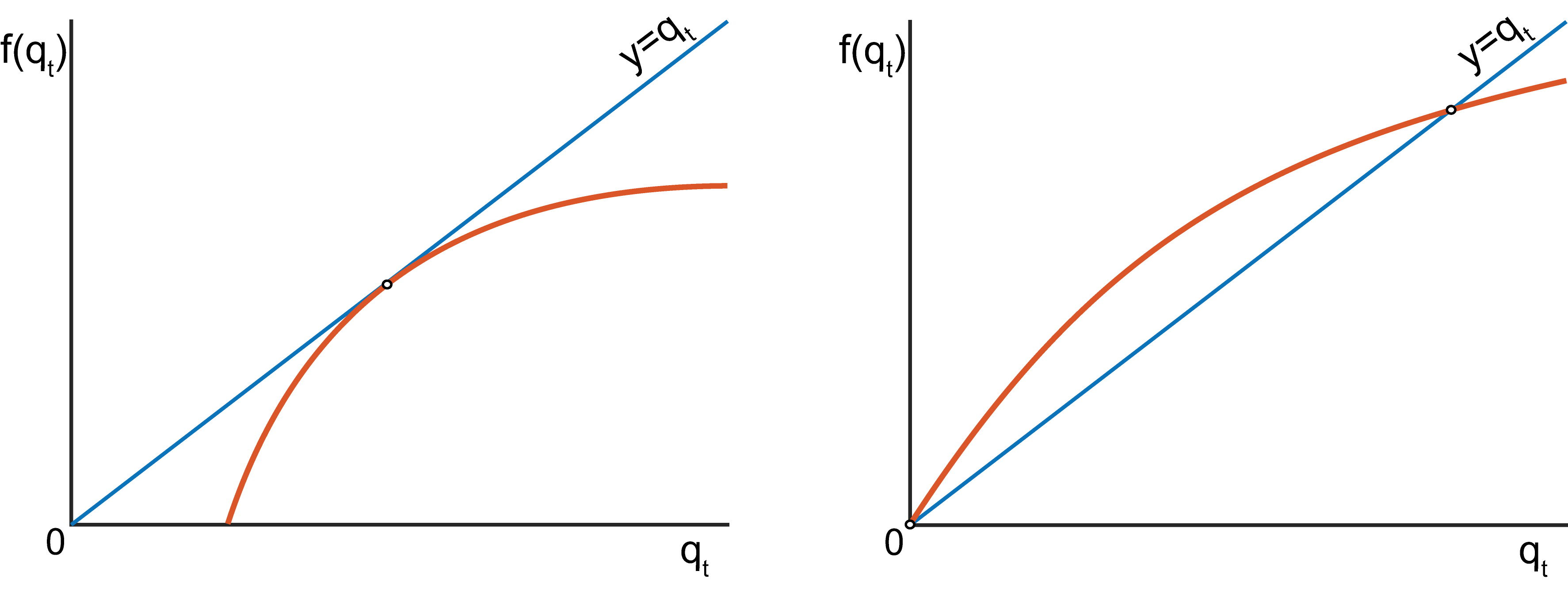}
\caption{Illustration to the proof of Proposition \ref{prop:no_spikes}.}\label{fig:proof}
\end{figure}
If the value of $p$ is such that the right derivative of
\[
f(q_t)=1-(1-p) ^ {zq_t}
\]
$q_t=0$  is less or equal to $1$ then strict concavity of $f(\cdot)$ implies that $f(q_t)<q_t$ for all $q_t\in(0,1]$. Hence (\ref{eq:1.41}) has only one fixed point, $q_t=0$. The corresponding condition is $-z\log(1-p)\leq 1$. This fixed point is attracting: $ \lim_ {t\rightarrow\infty} q_t=0$. If $-z\log(1-p)> 1$ then the trivial equilibrium $q_t=0$ becomes a repeller and the second fixed point $q_t=\tilde q$, $\tilde q \in (0,1)$ appears. At this point, the line $y=q$ and the curve $y=f(q)=1-(1-p) ^{zq}$ intersect transversely. Indeed, if this is not the case then there is a point $q'>0$ such that $1-(1-p) ^{zq'} = 0$ (see Fig. \ref{fig:proof}, left panel). The latter, however, is impossible as $(1-p)\in(0,1)$. Moreover, at the point of this intersection, the slope of the curve $y=f(q)=1-(1-p) ^{zq}$ is always strictly smaller than one (see Fig. \ref{fig:proof}, right panel). In order to see this, recall that the following must hold at $q=\tilde{q}$:
\[
\tilde{q}=f(\tilde{q})=\int_0^{\tilde{q}} \frac{df}{dq} (q) dq.
\]
If $df/dq (\tilde{q})\geq 1$ then strict concavity of $f$ implies that $df/dq (q) > 1$ for all $q\in(0,\tilde{q}]$. This, however, contradicts to that $\tilde{q}=f(\tilde{q})$. Hence the slope of the function $f(\cdot)$ at the second fixed point, $\tilde{q}$, is strictly smaller than one, and the fixed point  is locally exponentially stable.  $\square$

\subsection*{Proof of Proposition \ref{prop:energy_model}}

{\it Forward invariance}. Similar to the proof of Proposition \ref{prop:no_spikes}, we observe that $0<p\sigma(E_t,\underline{E})<1$ for all $E_t$. Hence  $a=(1-p\sigma(E_t,\underline{E}))^{z}\in(0,1)$ and $1-a^{q_t}\in[0,1]$ for all $q_t\in[0,1]$. This implies that $q_t\in [0,1]$ for all $t$ regardless of the values of $E_t$. Let $E_t\in [0, \overline{E}]$ and $E_t < r q_t$. In this case, $E_{t+1}=(1-\varepsilon)E_t + \varepsilon \overline{E}\in [0,\overline{E}]$. If $E_t\in [0, \overline{E}]$ and $E_t \geq r q_t$,  then
\[
 E_{t+1}=(1-\varepsilon)E_t + \varepsilon \overline{E} - r q_t \geq \varepsilon (\overline{E}-E_t) \geq 0
\]
\[
 E_{t+1}\leq (1-\varepsilon)\overline{E}+ \varepsilon \overline{E} - r q_t \leq  \overline{E}.
\]
Thus the domain $\{(q,E) \ | q\in [0,1], E\in [0, \overline{E}]\}$ is forward-invariant.

{\it Statement 1}. Observe, that $q^\ast_1=0$, $E^\ast_1=\overline{E}$ is always an equilibrium of (\ref{eq:mean_field_adaptive}). At this equilibrium, $\overline{E}=E^\ast_1\geq r q_1^\ast = 0$, and hence the Jacobian $J(q^\ast_1,E^\ast_1)$ of the right-hand side of (\ref{eq:mean_field_adaptive}) at this equilibrium is:
\[
J(0,\overline{E})=\left(\begin{array}{cc}
                        -z \log(1-p\sigma(\overline{E},\underline{E})) & 0\\
                        -r & 1-\varepsilon
                        \end{array}\right)
\]
It is clear that the eigenvalues of $J(0,\overline{E})$ are $\lambda_1=-z \log(1-p\sigma(\overline{E},\underline{E}))$, $\lambda_2=1-\varepsilon$. Therefore, the fixed point $(q_1^\ast,E_1^\ast)$ is an attractor when $-z \log(1-p\sigma(\overline{E},\underline{E}))<1$ and is a repeller when $-z \log(1-p\sigma(\overline{E},\underline{E}))>1$. This proves statement 1 of the proposition.

{\it Statements 2,3.} Let $(q^\ast,E^\ast)$ with $q^\ast\neq 0$ be another equilibrium of (\ref{eq:mean_field_adaptive}). All such equilibria of (\ref{eq:mean_field_adaptive}) must satisfy
\[
\begin{split}
q^\ast=& 1 - (1-p\sigma(E^\ast,\underline{E}))^{q^\ast z}\\
E^\ast=& (1-\varepsilon)E^\ast +\varepsilon\overline{E}-{rq^\ast}\mathcal{H}(E^\ast-r q^\ast).
\end{split}
\]
Depending on the sign of $E^\ast-r q^\ast$, the above system splits into the following two cases:
\begin{equation}\label{eq:mean_field_adaptive:eq1}
\left\{\begin{aligned}
q^\ast=& 1 - (1-p\sigma(E^\ast,\underline{E}))^{q^\ast z}\\
E^\ast=& \overline{E}-\frac{rq^\ast}{\varepsilon}, \ \mbox{if} \ E^\ast \geq r q^\ast,
\end{aligned}\right.
\
\left\{\begin{aligned}
q^\ast=& 1 - (1-p\sigma(E^\ast,\underline{E}))^{q^\ast z}\\
E^\ast=& \overline{E}, \ \mbox{if} \ E^\ast < r q^\ast.
\end{aligned}\right.
\end{equation}
Let $g(\cdot)=\sigma(\cdot,\underline{E})$. Note that the function $g(\cdot): \Real \rightarrow (0,1)$ is continuous and strictly increasing for $w>0$. Hence $g^{-1}(\cdot): (0,1)\rightarrow \Real$ exists, and is  continuous and  strictly increasing too. Moreover, points $(q^\ast,E^\ast)$ satisfying (\ref{eq:mean_field_adaptive:eq1}) should satisfy  conditions below (and vice-versa):
\begin{equation}\label{eq:mean_field_adaptive:eq2}
\left\{\begin{aligned}
E^\ast = & g^{-1}\left(\frac{1}{p}\left(1-(1-q^\ast)^{\frac{1}{q^\ast z}}\right)\right), \\
E^\ast=& \overline{E}-\frac{rq^\ast}{\varepsilon}, \ \mbox{if} \ E^\ast \geq r q^\ast,
\end{aligned}\right. \
\left\{\begin{aligned}
E^\ast = & g^{-1}\left(\frac{1}{p}\left(1-(1-q^\ast)^{\frac{1}{q^\ast z}}\right)\right), \\
E^\ast=& \overline{E}, \ \mbox{if} \ E^\ast < r q^\ast.
\end{aligned}\right.
\end{equation}

Consider the functions
\[
\begin{aligned}
h(q^\ast)=&g^{-1}\left(\frac{1}{p}\left(1-(1-q^\ast)^{\frac{1}{q^\ast z}}\right)\right),\\
f(q^\ast)=& (1-q^\ast)^{\frac{1}{q^\ast}}.
\end{aligned}
\]
Let $\Omega$ be the domain of the definition of the function $h(\cdot)$. If a solution of (\ref{eq:mean_field_adaptive:eq2}) exists with $q^\ast\in(0,1)$ then the intersection
\[
(0,\overline{q})=\Omega\cap (0,1)
\]
must be non-empty. The function $f(\cdot)$ is continuous on $(0,1)$, and its first derivative,
\[
\begin{split}
& \frac{d}{dq^\ast} f =(1-q^\ast)^{\frac{1}{q^\ast}}\left(-\frac{1}{{q^\ast}^2}\log (1-q^\ast)-\frac{1}{q^\ast(1-q^\ast)}\right)\\
&= -\frac{1}{{q^\ast}^2} (1-q^\ast)^{\frac{1}{q^\ast}}\left(\log(1-q^\ast)+\frac{q^\ast}{1-q^\ast}\right),
\end{split}
\]
is strictly negative in $(0,1)$.
%
This implies that the function $h(\cdot)$ is strictly increasing in $(0,\overline{q})$.

Consider the case when $E^\ast\geq r q^\ast$, (left system in (\ref{eq:mean_field_adaptive:eq2})). Given that $E^\ast=\overline{E}-rq^\ast/\varepsilon$, as a function of $q^\ast$, is strictly decreasing on $(0,1)$, and $h(\cdot)$ is strictly increasing on $(0,\overline{q})$, there must be at most one equilibrium of (\ref{eq:mean_field_adaptive}) in $(0,\min\{1,\overline{q}\})\subset(0,1)$. If such equilibrium exists then it must satisfy
\begin{equation}\label{eq:equilibrium_condition:1}
q^\ast=1-\left(1-p\sigma\left(\overline{E}-\frac{r}{\varepsilon}q^\ast,\underline{E}\right)\right)^{z q^\ast}
\end{equation}
for some $q^\ast\in(0,1)$. This, however,  is possible only if the the derivative of the right-hand side of (\ref{eq:equilibrium_condition:1})  at $q^\ast=0$ is larger than $1$. The corresponding condition is
\[
\begin{split}
&\frac{d}{dq^\ast}\left(1-\left(1-p\sigma\left(\overline{E}-\frac{r}{\varepsilon}q^\ast,\underline{E}\right)\right)^{z q^\ast}\right)= - \frac{d}{dq^\ast} \left(\left(1-p\sigma\left(\overline{E}-\frac{r}{\varepsilon}q^\ast,\underline{E}\right)\right)^{z q^\ast}\right)=\\
&- \left.\left(1-p\sigma\left(\overline{E}-\frac{r}{\varepsilon}q^\ast,\underline{E}\right)\right)^{z q^\ast}\right|_{q^\ast=0} \times\\
 & \left.\times\left[z \log\left(1-p\sigma\left(\overline{E}-\frac{r}{\varepsilon}q^\ast,\underline{E}\right)\right)+\frac{zq^\ast p}{1-p\sigma\left(\overline{E}-\frac{r}{\varepsilon}q^\ast,\underline{E}\right)} \frac{d}{dq^\ast} \sigma\left(\overline{E},\underline{E}\right)  \right] \right|_{q^\ast=0}\\
&=-z \log\left(1-p\sigma\left(\overline{E},\underline{E}\right)\right)>1.
\end{split}
\]

Consider the case when $E^\ast< r q^\ast$ (the right system in (\ref{eq:mean_field_adaptive:eq2})). Equilibria corresponding to this alternative must satisfy $q^\ast>\overline{E}/r$. This, however, is  possible only if $\overline{E}/r<1$. The alternative condition, $\overline{E}/r\geq 1$, therefore implies that $E^\ast\geq r q^\ast$. The latter observation completes the proof of statement 2.

Let $\overline{E}/r<1$ (or $E^\ast< r q^\ast$). Given that the function $h(\cdot)$ is strictly monotone on  $(0,\min\{1,\overline{q}\})$, only one equilibrium with $E^\ast=\overline{E}$ may exist. At this equilibrium,
\[
z \log\left(1-p\sigma\left(\overline{E},\underline{E}\right)\right) = \frac{\log(1-q^\ast)}{q^\ast}<-1
\]
and
\[
\left|\frac{d}{d q^\ast} (1-(1-p\sigma(\overline{E},\underline{E}))^{q^\ast z})\right| < 1.
\]
The latter inequality is due to that $1-(1-p\sigma(\overline{E},\underline{E}))^{q^\ast z}$ is strictly concave with respect to $q^\ast$ on $(0,1)$ (see the proof of Proposition \ref{prop:no_spikes}). Consider the Jacobian $J(q^\ast, \overline{E})$:
\[
J(q^\ast,\overline{E})=\left(\begin{array}{cc} \frac{d}{d q^\ast} (1-(1-p\sigma(\overline{E},\underline{E}))^{q^\ast z}) & [*]\\
                                0 & (1-\varepsilon)
                               \end{array}
                                 \right),
\]
where $[*]$ stands for the corresponding entry of the Jacobian matrix. The absolute values of its eigenvalues are clearly less than $1$, and hence the fixed point is a stable attractor. $\square$.



\begin{thebibliography}{99}

\bibitem{Abott:07}
L. Abbott, R. Rohrkemper (2007). A single growth model constructs critical avalanche networks, Prog Brain Res, 165, 9--13.

\bibitem{Bak:97}
P. Bak (1997). How Nature Works. The Science of Self-organized Criticality, Oxford Univ. Press.

\bibitem{beggs:03}
J. M. Beggs, D. Plenz (2003).  Neuronal avalanches in neocortical circuits, J Neurosci., 23, 11167--11177.

\bibitem{beggs:04}
J. M. Beggs, D. Plenz (2004). Neuronal avalanches are diverse and precise activity patterns that are stable for many hours in cortical slice cultures,  J Neurosci., 24, 5216--5229.

\bibitem{Cowan:10}
M. Benayoun, J.D. Cowan, W. van Drongelen, E. Wallace (2010). Avalanches in a stochastic model of spiking neurons, PLOS Computational Biology, 6(7), e1000846.

\bibitem{Bornholdt:00}
S. Bornholdt, T. Rohlf (2000). Topological Evolution of Dynamical
Networks: Global Criticality from Local Dynamics, Phys. Rev. Lett., 84, 26, 6114--6117, doi 10.1103.

\bibitem{Censi:2011}
F. Censi, A. Giuliani, P. Bartolini, G. Calcagnini (2011). A multiscale graph theoretical
approach to gene regulation networks: a case study in atrial fibrillation,
Biomedical Engineering, IEEE Transactions on, 58 (10), 2943–2946.


\bibitem{Christensen:98}
K. Christensen , R. Donangelo, B. Koiller , K. Sneppen (1998). Evolution of Random Networks, PhysRevLett. 81.2380, doi 10.1103.

\bibitem{Dong:2009}  X.-X. Dong, Y. Wang, and Z.-H. Qin (2009). Molecular mechanisms of excitotoxicity and their relevance to pathogenesis of neurodegenerative diseases, Acta Pharmacologica Sinica, (30) 4, 379–387.

\bibitem{Beggs:2012} N. Friedman, S. Ito, B.A. Brinkman, M. Shimono, R.L. DeVille, K.A. Dahmen, J.M. Beggs, T.C. Butler (2012). Universal critical dynamics in high resolution neuronal avalanche data. Physical review letters, 108(20), p.208102.

\bibitem{Gavello:2012}

D. Gavello, J. Rojo-Ruiz, A. Marcantoni, C. Franchino, E. Carbone, V, Carabelli (2012). Leptin Counteracts the Hypoxia-Induced Inhibition of Spontaneously Firing Hippocampal Neurons: A Microelectrode Array Study,  PLOS ONE, 10.1371/journal.pone.0041530.

\bibitem{gong:04}
P. Gong, C. van Leeuwen (2004). Evolution to a small-world network with chaotic units, Europhysics Letters, 67(2), 328.

\bibitem{gong:07}
P. Gong, C. van Leeuwen (2007). Dynamically maintained spike timing sequences in networks of pulse-coupled oscillators with delays, Physical Review Letters, 98(4), 048104.

\bibitem{Gorban:1981} A. N. Gorban, V. M. Cheresiz (1981). Slow Relaxations of Dynamic-Systems and Bifurcations of Omega-Limit Sets. DOKLADY AKADEMII NAUK SSSR, 261(5), 1050-1054 {\it (communicated by S.L. Sobolev)}.

\bibitem{Gorban:2004} A.N. Gorban (2004). Singularities of Transition Processes in Dynamical Systems: Qualitative Theory of Critical Delays. Electronic Journal of Differential Equations, 2004.

\bibitem{Gorban:09}
A.N. Gorban, E.V. Smirnova, T.A. Tyukina (2009). General Laws of Adaptation to Environmental Factors: from Ecological Stress to Financial Crisis. Math. Model. Nat. Phenom., 4(6), 1-53 (2009).

\bibitem{Gorban:10}
A.N. Gorban, E.V. Smirnova, T.A. Tyukina (2010). Correlations, risk and crisis: From physiology to finance, Physica A, volume 389, 16, 3193--3217.

\bibitem{Gorban:15}
A.N. Gorban, T.A. Tyukina, E.V. Smirnova, L.I. Pokidysheva (2016). Evolution of adaptation mechanisms: adaptation energy, stress, and oscillating death, J. Theor. Biol., 405, 127--139, http://dx.doi.org/10.1016/j.jtbi.2015.12.017, http://arxiv.org/pdf/1512.03949.pdf.

\bibitem{Gritsun:12}
T. A. Gritsun, J. le Feber, W. L. C. Rutten (2012). Growth Dynamics Explain the Development of Spatiotemporal Burst Activity of Young Cultured Neuronal Networks in Detail, PLOS ONE. 7(9), e43352, doi: 10.1371/journal.pone.0043352.

\bibitem{Iudin:13}
D.I. Iudin, E.V. Koposov (2013). Fractals: as simple as complex, NiSOC.

\bibitem{Iudin:12}
D.I. Iudin, Ya.D. Sergeyev,  M. Hayakawa (2012). Interpretation of percolation in terms of infinity computationts. Applied Mathematics and Computation. 218(16), 8099--8111.

\bibitem{Iudin2:15}
D.I. Iudin, Ya.D. Sergeyev,  M. Hayakawa (2015). Infinity computations in cellular automaton forest-fire model., 20(3), 861--870.

\bibitem{Iudin:15}
F.D. Iudin, D.I. Iudin, V.B. Kazantsev (2015). Percolation treshold in active neural networks with adaptive geometry, JETP Lett., 101 (4).

\bibitem{Izhikevich:04}
E.M. Izhikevich, J.A. Gally, G.M. Edelman (2004). Spike-Timing Dynamics of Neuronal Groups, Cereb Cortex, 14, 933--944.

\bibitem{Izhikevich:06}
 E.M. Izhikevich (2006). Polychronization: Computation With Spikes, Neural Comput., 18, 245--282.

\bibitem{Levina:2007} A. Levina, J.M. Hermann, T. Geisel (2007). Dynamical synapses causing self-organized criticality in neural networks, Nature Physics, 3, 857--860.

\bibitem{Masquelier:13}
T. Masquelier, G. Deco (2013). Network Bursting Dynamics in Excitatory Cortical Neuron Cultures Results from the Combination of Different Adaptive Mechanisms, PLoS ONE 8(10): e75824. doi:10.1371/journal.pone.0075824.

\bibitem{Mohapatra:2009}
D. P. Mohapatra, H. Misonou, P. Sheng-Jun, J. E. Held, D. J. Surmeier, J. S. Trimmer (2009). Regulation of intrinsic excitability in hippocampal
neurons by activity-dependent modulation of the KV2.1 potassium channel, Channels, (3) 1, 46-56, DOI: 10.4161/chan.3.1.7655.

\bibitem {Chiappalone:08}
V. Pasqualea, P. Massobriob, L.L. Bolognaa, M. Chiappalone, S. Martinoia (2008). Self-organization and neuronal avalanches in networks of dissociated cortical neurons, Neuroscience, 153 (4), 1354--1369.

\bibitem{Meldrum:2000} B. S. Meldrum (2000). Glutamate as a neurotransmitter in the brain: review of physiology and pathology, Journal of Nutrition, (153) 4, 1007S–-1015S.

\bibitem{Nguyen:2011} D. Nguyen, M. V. Alavi, K.-Y. Kim,  T. Kang, R. T. Scott, Y. H. Noh, J. D. Lindsey, B. Wissinger, M. H. Ellisman, R. N. Weinreb, G. A. Perkins, and W.-K. Ju (2011).  A new vicious cycle involving glutamate excitotoxicity, oxidative stress and mitochondrial dynamics, Cell Death and Disease, (2)12, article e240.



\bibitem{Pimashkin:11}
A.S. Pimashkin, I.A. Kastalskiy, A.Yu. Simonov, E.A. Koryagina, S.A. Korotchenko, I.V. Mukhina , V.B. Kazantsev (2011). Spiking signatures of spontaneous activity bursts in hippocampal cultures. Frontiers in Computational Neuroscience, 5(46).

\bibitem{Ziff:07}
John A. Quintanilla, R. M. Ziff (2007). Near symmetry of percolation thresholds of fully penetrable disks with two different radii, Phys, Rev. E 76 (5), 051115.

\bibitem{Selye:38a} H. Selye, 1938. Adaptation energy, Nature 141 (3577), 926.

\bibitem{Selye:38b} H. Selye, 1938. Experimental evidence supporting the conception of ``adaptation energy'', Am. J. Physiol. 123, 758--765.




\bibitem{Kirkpatrick:71}
K. S. Shante, S. Kirkpatrick. An introduction to percolation theory. Advances in Physics 20 (85): 325--357.



\bibitem{Tetzlaff:10}
C. Tetzlaff, S. Okujeni, U. Egert, F. W\"{o}rg\"{o}tter, M. Butz (2010). Self-Organized Criticality in Developing Neuronal Networks. PLOS Comp. Bio. 6(12): e1001013.

\bibitem{Tanya:GitHub} Supplementary materials for ``Simple model of complex dynamics of activity patterns in
developing networks of neuronal cultures''. {https://github.com/tt51Storage/Simple-model-of-complex-dynamics-in-neuronal-cultures}.

\bibitem{Markram:00}
M. Tsodyks, A. Uziel, H. Markram (2000). Synchrony generation in recurrent networks with frequency-dependent synapses, J. Neuroscience, 20, 825--835.


\bibitem{Vedunova:13}
M. Vedunova, T.  Sakharnova, E. Mitroshina, M. Perminova, A. Pimashkin, Yu. Zakharov, A. Dityatev, I. Mukhina (2013). Seizure-like activity in hyaluronidase-treated dissociated hippocampal cultures, Frontiers in Cellular Neuroscience, 149 (8), doi: 10.3389/fncel.2013.00149.

\bibitem{Vedunova:14} M. V. Vedunova,  E. V. Mitroshina, T. A. Sakharnova, M. Yu. Bobrov, V. V. Bezuglov, L. G. Khaspekov, I. V. Mukhina (2014).  Effect of N-Arachidonoyl Dopamine on Activity of Neuronal Network in Primary Hippocampus Culture upon Hypoxia Modelling, Bulletin of Experimental Biology and Medicine 156 (4), 461-464.

\bibitem{Vedunova:15} M.V. Vedunova, T.A. Mishchenko, E.V. Mitroshina, I.V. Mukhina (2015). TrkB-mediated neuroprotective and antihypoxic properties of Brain-derived neurotrophic factor, Oxidative Med. Cell. Longevity, 2015,  p. 9, 10.1155/2015/453901.


\bibitem{Zhouand:2014} Zhouand, N.C. Danbolt (2014). Glutamate as a neurotransmitter in the healthy brain, Journal of Neural Transmission, 121 (8) 799–817.

\end{thebibliography}
\end{document}